\documentclass[%
twocolumn,
nofootinbib,
amsmath,amssymb,
aps,
prd,
floatfix,
]{revtex4-2}

\usepackage{graphicx}
\usepackage{dcolumn}
\usepackage{bm}
\usepackage{algorithm}
\usepackage{algpseudocode}
\usepackage{algorithmicx}
\usepackage{acronym}
\usepackage[normalem]{ulem}
\usepackage{multirow}
\usepackage[breaklinks, plainpages=false, colorlinks=true, 
anchorcolor=cyan, linkcolor=magenta, citecolor=cyan, urlcolor=purple, bookmarks=false]{hyperref}
\usepackage{xcolor} 

\DeclareMathOperator\erf{erf}

\def\lm{{\ell m}}
\def\mpsi{\psi_4^{\ell m}}
\def\mpsitt{\psi_4^{22}}
\def\mnews{N_4^{\ell m}}

\newcommand{\gra}{\texttt{GR-Athena++}}
\newcommand{\pit}{\texttt{PITTNull}}
\newcommand{\eq}[1]{Eq.~(\ref{#1})}
\newcommand{\fig}[1]{Fig.~\ref{#1}}
\newcommand{\app}[1]{App.~\ref{#1}}
\newcommand{\rf}[1]{Ref.~\cite{#1}}

\newcommand{\mrm}[1]{\mathrm{#1}}
\newcommand{\code}[1]{\texttt{#1}}
\newcommand{\qt}[1]{``#1''}
\newcommand{\mb}{\texttt{MeshBlock}}
\newcommand{\mbs}{\texttt{MeshBlocks}}

\newcommand\be{\begin{equation}}
\newcommand\ee{\end{equation}}

\newacro{BNS}{binary neutron star}
\newacro{BBH}{binary black hole}
\newacro{NS}{neutron star}
\newacro{BH}{black hole}
\newacro{CCE}{Cauchy characteristic extraction}
\newacro{WE}{Whittaker-Eilers}
\newacro{GW}{gravitational wave}
\newacro{FRE}{finite radius extraction}
\newacro{FFI}{fixed frequency integration}
\newacro{gw}[GW]{gravitational wave}
\newacro{xg}[XG]{next-generation}
\newacro{imr}[IMR]{inspiral-merger-ringdown}
\newacro{nr}[NR]{numerical relativity}
\newacro{pn}[PN]{post-Newtonian}
\newacro{pe}[PE]{parameter estimation}
\newacro{psd}[PSD]{power spectral density}
\newacro{et}[ET]{Einstein Telescope}
\newacro{eob}[EOB]{effective-one-body}
\newacro{ce}[CE]{Cosmic Explorer}
\newacro{imbh}[IMBH]{intermediate-mass black hole}

\begin{document}

\title{Binary Black Hole Waveforms from High-Resolution GR-Athena++ Simulations}

\author{Alireza \surname{Rashti}$^{1,2,3}$}
\author{Rossella \surname{Gamba}$^{1,2,3,4}$}
\author{Koustav \surname{Chandra}$^{1,2,3}$}
\author{David \surname{Radice}$^{1,2,3}$}
\author{Boris \surname{Daszuta}$^{5}$}
\author{William \surname{Cook}$^{5}$}
\author{Sebastiano \surname{Bernuzzi}$^{5}$}

\affiliation{$^{1}$
    Institute for Gravitation and the Cosmos,
    The Pennsylvania State University,
    University Park, PA 16802, USA}
\affiliation{$^{2}$
    Department of Physics,
    The Pennsylvania State University,
    University Park, PA 16802, USA}
\affiliation{$^{3}$
    Department of Astronomy \& Astrophysics,
    The Pennsylvania State University,
    University Park, PA 16802, USA}
\affiliation{${}^4$
		Department of Physics,
		University of California,
		Berkeley, CA 94720, USA}
\affiliation{${}^5$
		Theoretisch-Physikalisches Institut, 
		Friedrich-Schiller-Universit{\"a}t Jena,
		07743, Jena, Germany}

\begin{abstract}

The detection and subsequent inference of binary black hole signals rely 
heavily on the accuracy of the waveform model employed. 
In the highly non-linear, dynamic, and strong-field regime near merger, 
these waveforms can only be accurately modeled through numerical relativity 
simulations. Considering the precision requirements of next-generation gravitational 
wave observatories, we present in this paper high-resolution simulations of 
four non-spinning quasi-circular binary black hole systems with 
mass ratios of 1, 2, 3, and 4, conducted using the \gra{} code.
We extract waveforms from these simulations using both finite radius and 
\ac{CCE} methods. 
Additionally, we provide a comprehensive error analysis 
to evaluate the accuracy and convergence of the waveforms.
Our self-mismatch study shows that the (2, 2) mode of the \ac{CCE} strains,
for the world tube extraction radius of $R=50$,
reaches the level of $\sim 10^{-12}$ mismatch for mass ratios of 1, 2, 3, 
and $\sim 10^{-11}$ mismatch for the mass ratio of 4.
However, when larger extraction radii are considered or when more modes are included
the mismatches increase.
These results highlight both the promise and limitations of current simulations in achieving the 
precision required for upcoming detectors such as LISA, Cosmic Explorer, and Einstein Telescope.
The waveforms are publicly available on 
ScholarSphere~\cite{GRAthena:BHBH:0001,GRAthena:BHBH:0002,GRAthena:BHBH:0003,GRAthena:BHBH:0004},
and represent the first set of waveforms of the new \gra{} catalog.

\end{abstract}


\acresetall

\maketitle

\section{Introduction}
\label{sec:introduction}

The field of \ac{nr} achieved a major milestone in 2005 by successfully 
simulating \ac{BBH} systems with comparable masses through the stages of inspiral, merger, and 
ringdown~\cite{Pretorius:2005gq, Baker:2006ha, Campanelli:2005dd}. 
This accomplishment resulted from decades of work, beginning with Hahn and Lindquist in the 1960s~\cite{Hahn:1964a}, 
and involving contributions from numerous groups over the years. Today, several advanced codes, built on diverse numerical 
techniques and formulations, can robustly evolve \ac{BBH} mergers and extract the resulting 
\acp{gw}~\cite{Campanelli:2005dd, Scheel:2006gg, Sperhake:2006cy, Bruegmann:2006ulg, Brugmann:2008zz, Szilagyi:2009qz, Thierfelder:2011yi, Loffler:2011ay, Babiuc:2010ze, Hilditch:2015aba,Bugner:2015gqa,Clough:2015sqa, Kidder:2016hev,Daszuta:2021ecf}.

\ac{nr} simulations of the late-inspiral, merger, and ringdown phases of \acp{BBH} play a critical role 
in constructing and validating accurate waveform models used in \ac{gw} data analysis. 
These models are essential for tasks such as signal detection and parameter estimation. In particular, \ac{nr} 
simulations serve as the foundation for creating semi-analytic waveform models, including the 
\ac{eob}~\citep{Buonanno:1998gg, Buonanno:2000ef, Ramos-Buades:2021adz, Pompili:2023tna, Ramos-Buades:2023ehm, Chiaramello:2020ehz, Akcay:2020qrj, Gamba:2021ydi, Nagar:2023zxh, Nagar:2024dzj, Nagar:2024oyk} and phenomenological \ac{imr} models~\citep{Ajith:2007qp, Ajith:2007kx, Ajith:2009bn, Santamaria:2010yb, Husa:2015iqa, Khan:2015jqa,Pratten:2020fqn,Estelles:2020osj,Estelles:2021gvs,Hamilton:2021pkf,London:2017bcn,Garcia-Quiros:2020qpx,Khan:2019kot,Hannam:2013oca,Schmidt:2014iyl,Khan:2018fmp,Pratten:2020ceb}, 
as well as for developing NR-based interpolants like the surrogate models~\citep{Blackman:2014maa, Blackman:2017dfb, Varma:2018mmi, Varma:2019csw, Williams:2019vub}. 
Additionally, \ac{nr} simulations have been used to assess the sensitivity of searches to binary populations 
and to establish upper limits on merger rate density, particularly for \ac{imbh} 
binaries~\citep{Chandra:2020ccy, LIGOScientific:2014oec, Chandra:2021xvs}.

Therefore, the accuracy of these \ac{nr} simulations is a critical factor that constrains the 
precision of the waveform models used in \ac{gw} analysis. 
As detector sensitivity continues to improve, with current instruments like Advanced LIGO~\citep{LIGOScientific:2014pky} 
reaching new levels in their fourth observing run, and with the advent of \ac{xg} \ac{gw} detectors such as the 
\ac{ce}~\citep{Reitze:2019iox, Evans:2023euw} and \ac{et}~\citep{Punturo:2010zz, Maggiore:2019uih}, 
which are anticipated to be at least ten times more sensitive, the demand for even more accurate waveform 
models becomes increasingly urgent. Future space-based missions, such as 
LISA~\citep{amaro2017laser}, TianQin~\citep{TianQin:2015yph}, Taiji~\citep{TaijiScientific:2021qgx}, 
DECIGO~\citep{Kawamura:2006up}, and the Lunar Gravitational-Wave Antenna (LGWA)~\citep{Ajith:2024mie}
will further heighten this need. 

Although quantifying the exact degree of accuracy required for upcoming detectors is challenging, 
numerous studies have shown that the precision of current state-of-the-art waveforms must be significantly enhanced
to avoid biased source parameter estimates~\cite{Lindblom:2008cm, Kapil:2024zdn,Dhani:2024jja,Chandra:2024dhf}. 
Reference~\cite{Purrer:2019jcp} studied these accuracy requirements for \ac{BBH} systems observed by
\ac{et} and \ac{ce}, concluding that simulations must improve by approximately 
one or more orders of magnitude to meet the demands of the \ac{xg} detectors.
Similarly, \rf{Jan:2023raq} and \rf{Ferguson:2020xnm} recently evaluated the impact of finite grid resolution on \ac{nr}
waveform precision, emphasizing the need for higher-resolution \ac{nr} simulations, especially for \ac{ce} and LISA. 

Beyond studies assessing the accuracy improvement of simulations from the same catalogs, cross-comparisons
between simulations obtained with different codes and techniques~\cite{Ajith:2012az, LIGOScientific:2014oec, Hinder:2013oqa} 
also play a crucial role in validating the robustness of the resulting waveforms.
To this end, the production of new, high-resolution \ac{nr} simulations is crucial.

In this work, we use \gra{}~\cite{Daszuta:2021ecf,Rashti:2023wfe} to conduct 
high-resolution dynamical evolutions of non-spinning, quasi-circular \ac{BBH} 
systems with varying mass ratios. 
The total computational cost of these \ac{BBH} simulations is $\sim 155$ million
core-hours.
\textit{They constitute the first waveforms in the
\gra{}~catalog~\cite{GRAthena:BHBH:0001,GRAthena:BHBH:0002,GRAthena:BHBH:0003,GRAthena:BHBH:0004}},  
simulated with the aim of meeting the precision requirements of future \ac{gw} detectors. 
Their accuracy is thoroughly assessed through a series of tests, 
including direct comparisons with selected waveforms from the SXS catalog,
providing robust cross-validation with prior results.

The rest of this paper is structured as follows. In Sec.~\ref{sec:numerical_method}, we provide
a detailed overview of the numerical methods, including grid configurations, 
\ac{gw} extraction techniques, and error estimation models. 
In Sec.~\ref{sec:results} we present our simulations, showcasing
the waveforms for different modes along with the trajectories of the \ac{BBH}
systems for each mass ratio.
Sec.~\ref{sec:results} is dedicated to evaluating the quality of the waveforms
through a series of convergence studies and cross-validation against waveforms from the SXS catalog.
Specifically, we analyse the convergence of \ac{CCE} strains and estimate their
errors contributed from different parameters such as the resolution and
world tube extraction radius. Comparisons against SXS are performed
via time-domain alignment as well as frequency-domain mismatch calculations, considering
both single-mode and multi-mode waveforms.
Finally, Sec.~\ref{sec:discussion} concludes our work with an overview of our results
and reflections on future research directions.
Additional material for further examination is provided in the appendices.

Throughout the paper, we use geometrized units where $G = c = 1$, with $G$ representing 
the gravitational constant and $c$ the speed of light.

\section{Numerical simulations}
\label{sec:numerical_method}
%
\begin{table*}[hbt!]
\begin{center}
\begin{tabular}{|c|c|c|c|c|c|}
\hline
$q$ & $D/M$ & $10^{4}\times |P_x/M|$ & $10^{2} \times |P_y/M|$ & %
$N$ & $ 10^{3}\times \delta x/M $ \\
\hline
1 & 12 & 4.681 & 8.507 & %
128, 192, 256, 320, 384 & 19.65, 13.10, 9.83, 7.86, 6.55 \\
2 & 10 & 7.948 & 8.560 & %
128, 192, 256, 320, 384 & 5.981, 3.988, 2.991, 2.393, 1.994 \\
3 & 10 & 4.968 & 7.237 & %
128, 192, 256, 320, 384 & 6.775, 4.517, 3.387, 2.710, 2.258 \\
4 & 10 & 4.211 & 6.188 & %
128, 192, 256, 320 &  3.723, 2.482, 1.862, 1.489 \\
\hline
\end{tabular}
\end{center}
\caption{%
Parameters of \ac{BBH} simulations.
$q$ denotes the mass ratio, and $D$ is the distance between two \acp{BBH}.
$|P_x/M|$ and $|P_y/M|$ respectively denote $x$ and $y$ momentum components of
the \acp{BH} in the \code{TwoPunctures} initial data code. 
The total mass of the system $M = 1$ in all cases.
Each mesh block contains $16^{3}$ grid points. 
The mesh resolution $N$ specifies the number 
of grid points on the root grid in each direction while the grid space,
$\delta x$, 
indicates the resolutions at the finest level.
}
\label{tab:bbh_runs}
\end{table*}
\begin{figure*}[hbt!]
\centering
\includegraphics[width=0.8\linewidth,clip=true]{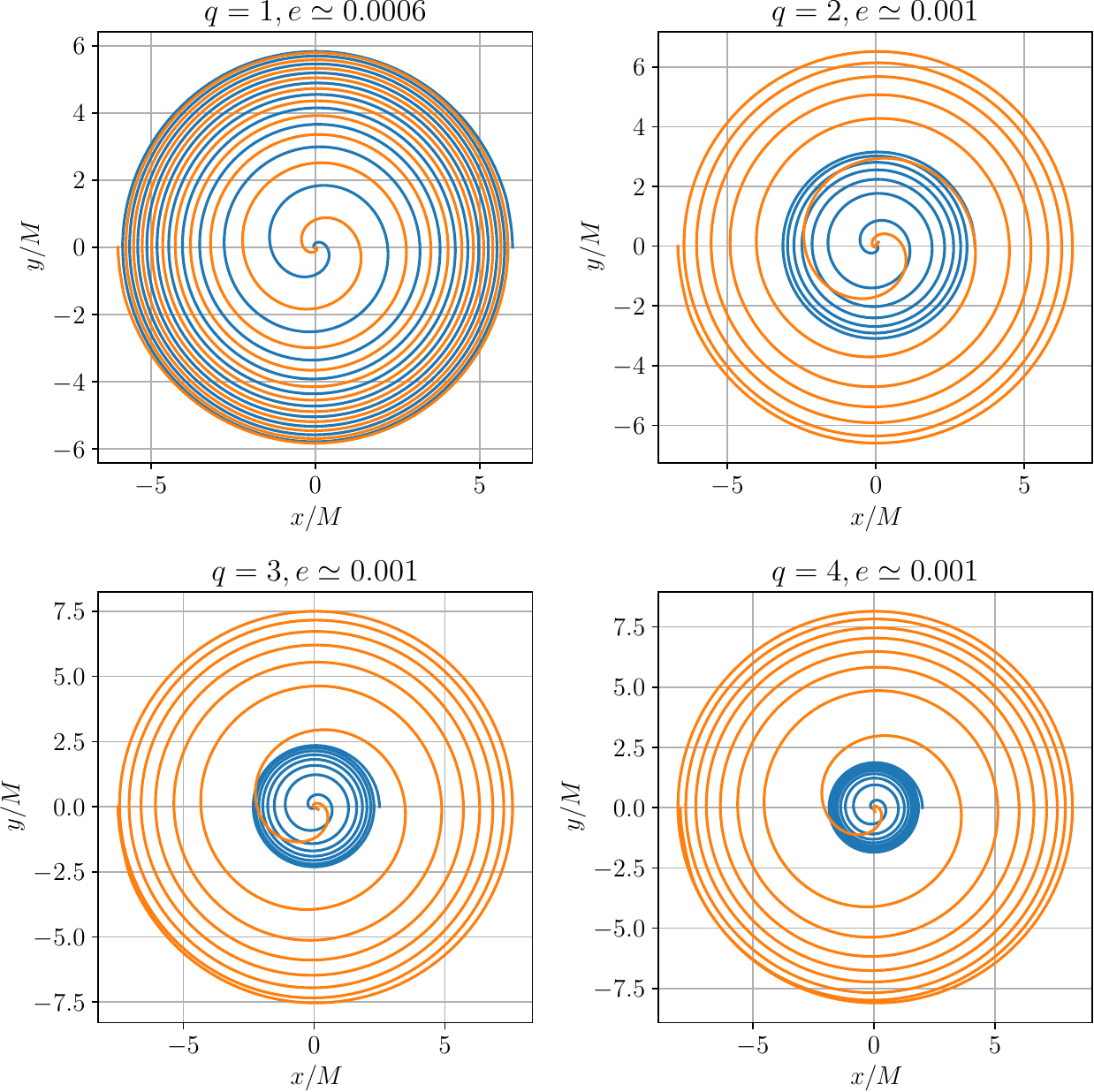}
\caption{%
Quasi-circular orbits of \ac{BBH} simulations. The orbits are drawn by
tracking the coordinate of each puncture. The mass ratio and
eccentricity of each run is specified. The eccentricity~($e$) approximated 
based on a fit to the coordinate distance for the first few orbits of the
run, see the text.
}
\label{fig:orbits}
\end{figure*}
\begin{figure*}[hbt!]
\centering
\includegraphics[width=1\linewidth,clip=true]{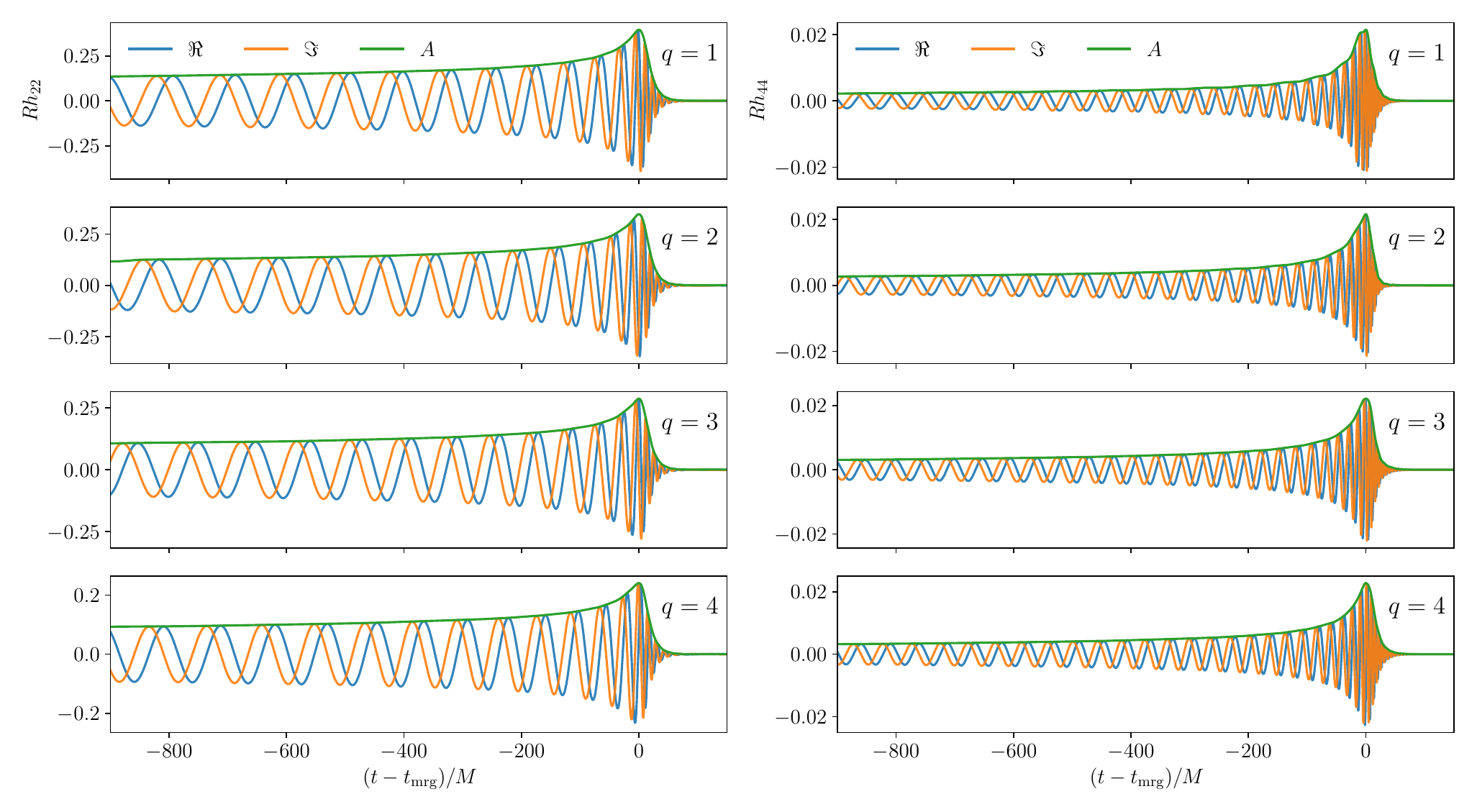}
\caption{%
This figure shows the real and the imaginary parts of $(2,2)$~[left] and
$(4,4)$~[right] multipoles of the \ac{CCE} strains 
for our simulations. We also show the amplitude evolution of these modes.
$t_{\mrm {mrg}}$ is defined as the peak amplitude time of the $(2,2)$ mode.
}
\label{fig:h_catalog}
\end{figure*}
\begin{figure*}[hbt!]
\centering
\includegraphics[width=1\linewidth,clip=true]{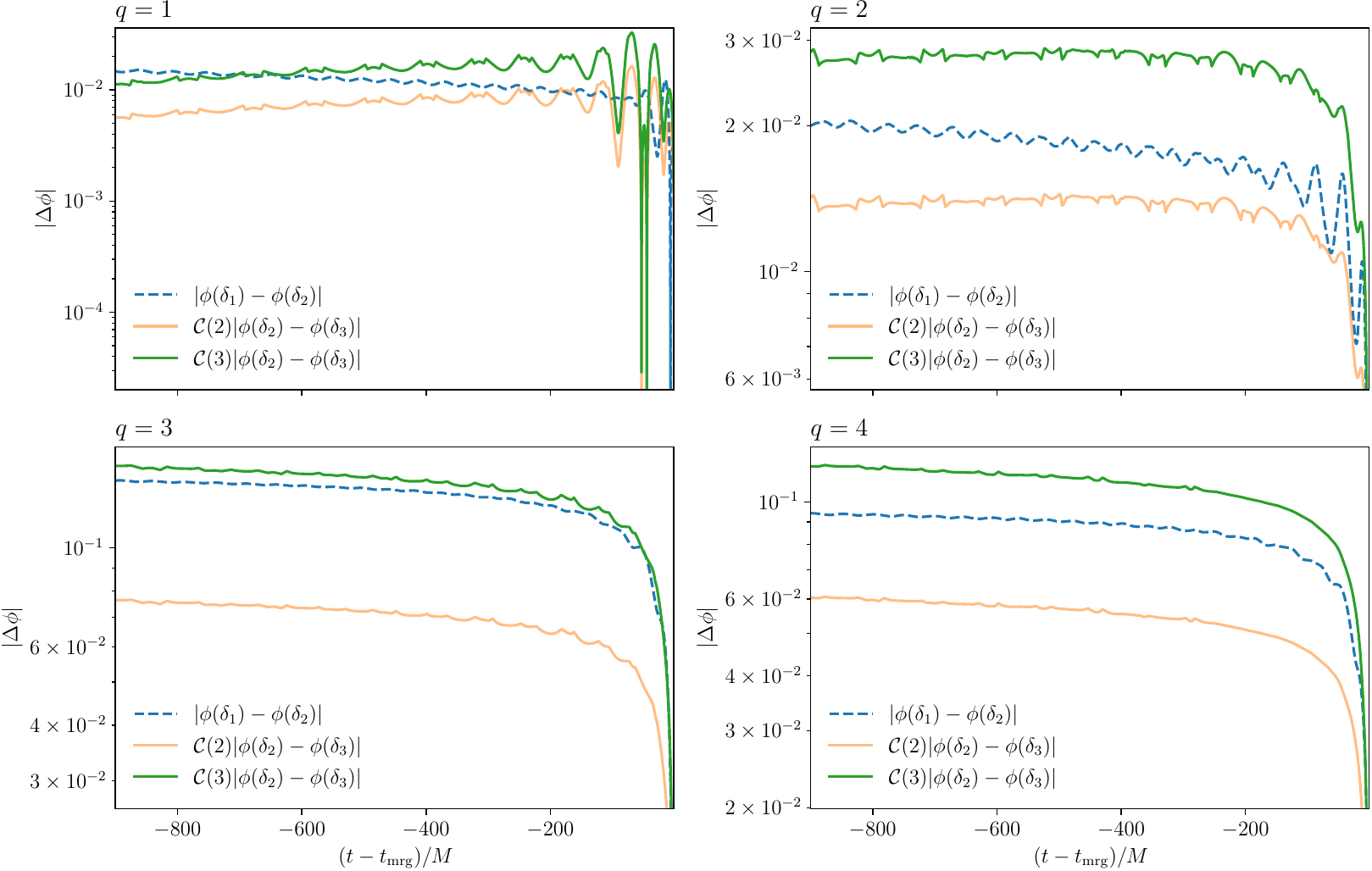}
\caption{%
Convergence of \ac{CCE} strains, (2,2) modes.
Here, we have  $\delta_{i}^{-1} \in [56, 112, 224]$ for $i = 1,2,3$.
The strains are computed from $\mpsi$.
The boundary condition for \pit{} is obtained by the highest resolution from the 
table~(\ref{tab:bbh_runs}) and the radius of world tube extraction is $50$.
The strains show $\gtrsim 2$nd order of convergence for $q=$ 1, 2, and 4 runs, 
while the strain of $q=3$ run exhibits $\sim 3$rd order of convergence.
Nevertheless, in our error analyses, we choose 2nd order convergence rate for all strains 
with respect to the \pit's resolution, expected from the \pit{} code~\cite{Babiuc:2010ze}.
}
\label{fig:conv_test_cce_h_from_psi4}
\end{figure*}
\begin{figure*}[hbt!]
\centering
\includegraphics[width=1\linewidth,clip=true]{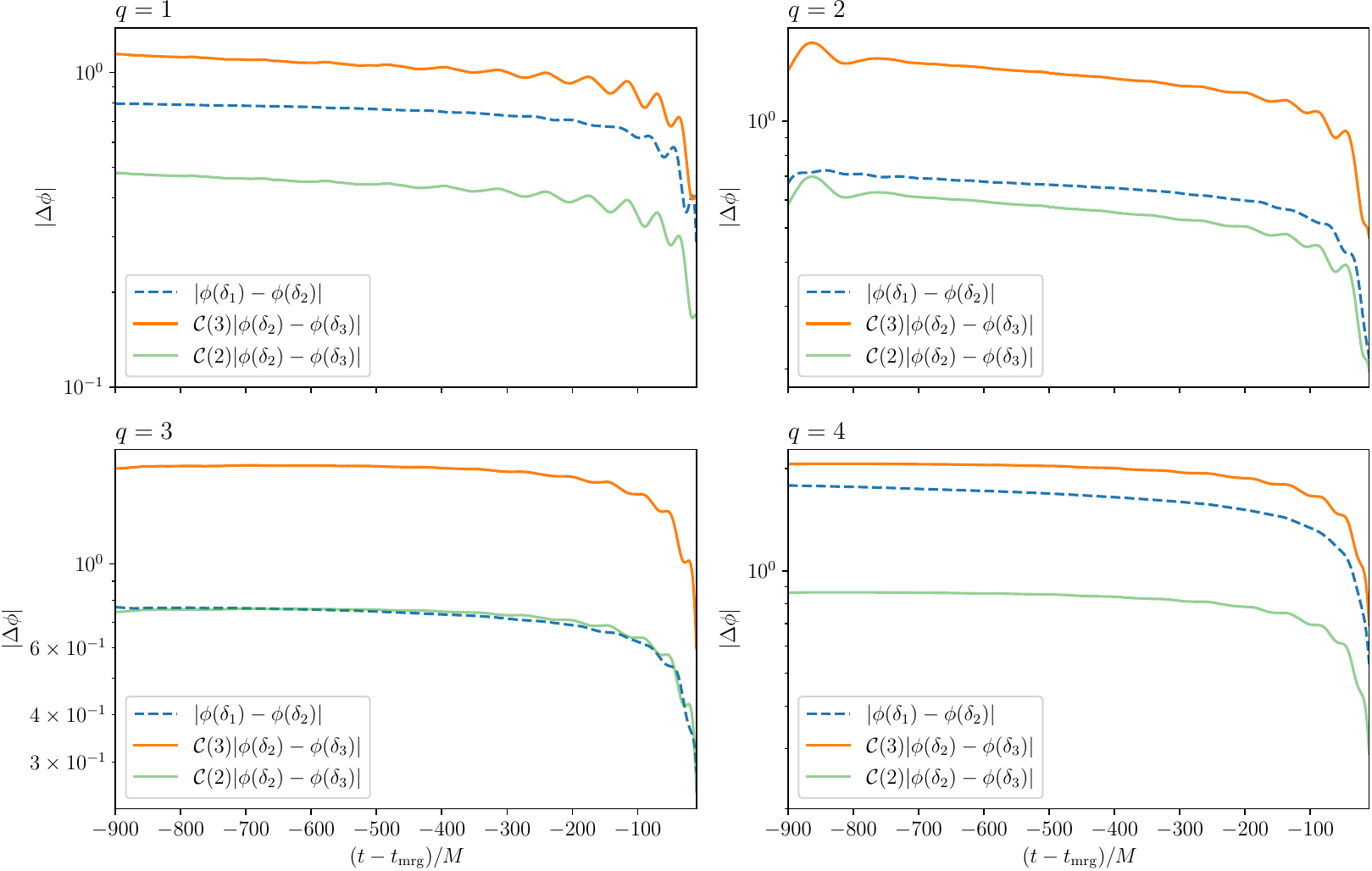}
\caption{%
Convergence of \ac{CCE} strains, (2,2) modes 
with respect to the \pit's world tube extraction radius.
Here, we have  $\delta_{i}^{-1} \in [20, 50, 100]$ for $i = 1,2,3$.
The strains are computed from $\mpsi$.
The boundary condition for \pit{} is obtained by the highest resolution from the 
table~(\ref{tab:bbh_runs}) and it is run using the highest resolution of the
\pit{} code, see the text.
}
\label{fig:conv_test_rex_cce_h_from_psi4}
\end{figure*}
\begin{figure*}[hbt!]
\centering
\includegraphics[width=1\linewidth,clip=true]{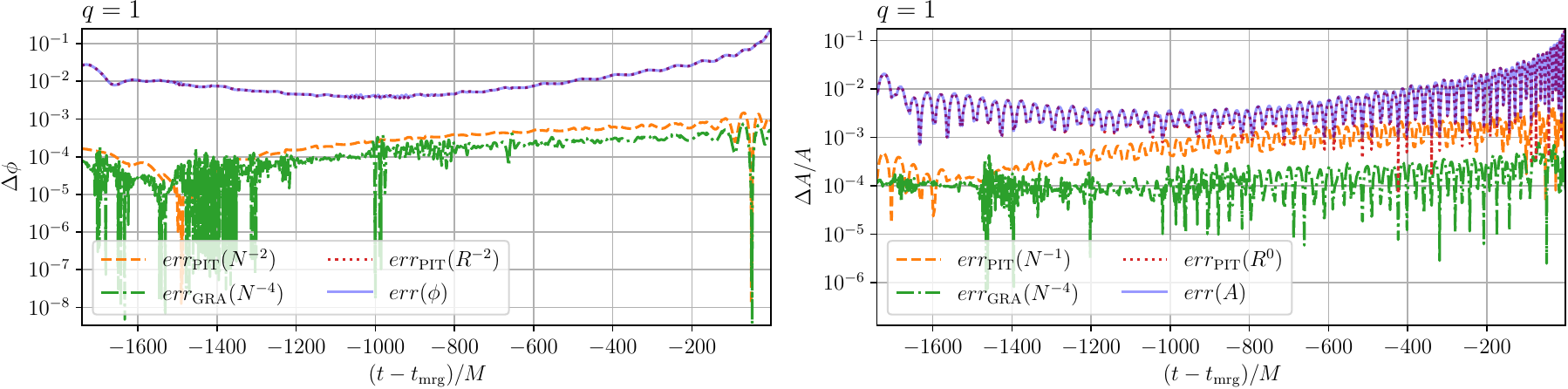}
\includegraphics[width=1\linewidth,clip=true]{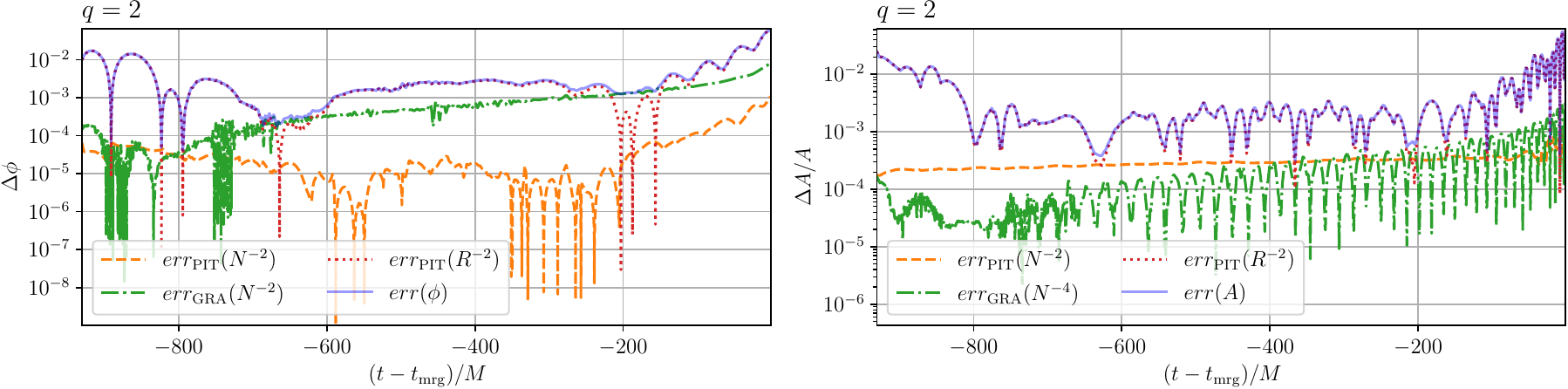}
\includegraphics[width=1\linewidth,clip=true]{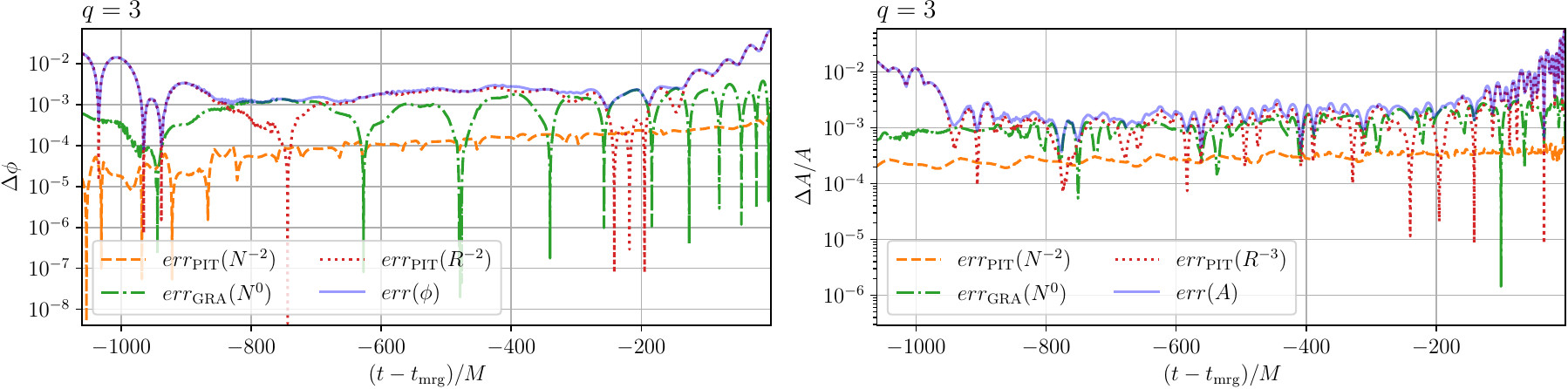}
\includegraphics[width=1\linewidth,clip=true]{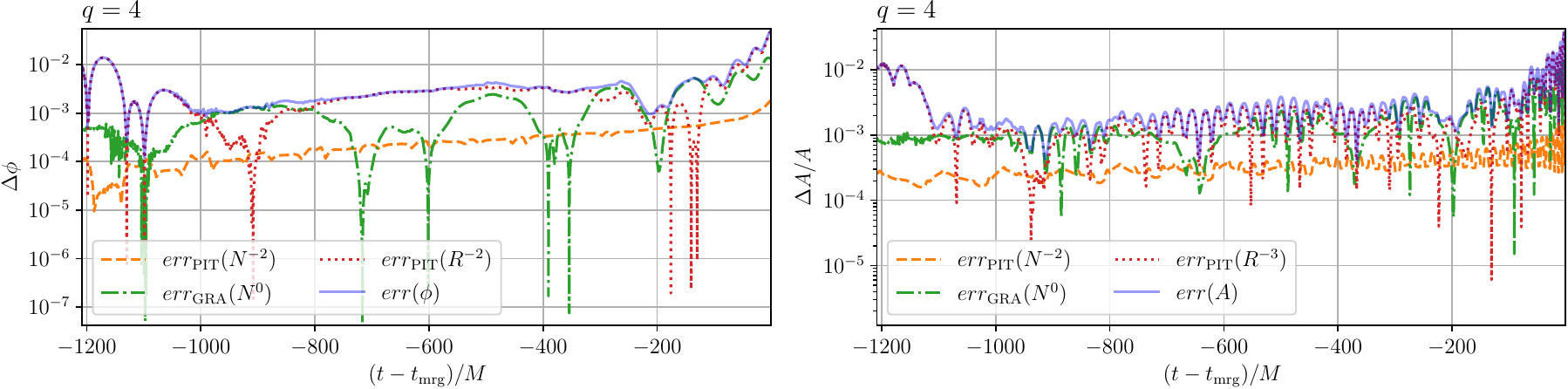}
\caption{%
Sum of errors for the $(2,2)$ mode of \ac{CCE} strains~(computed from $\mpsi$).
$err_{\mrm{PIT}}(N^{-p})$ denotes the magnitude of the leading order error
term, \eq{eq:trunc_error},
at the highest resolution of \pit{} run, and $p$ is the convergence order.
Similarly, $err_{\mrm{PIT}}(R^{-p})$ denotes the error due to world tube
extraction radius at $R=50$; 
$err_{\mrm{GRA}}(N^{-p})$ marks the error pertinent 
to \gra{}'s resolutions, at the highest available resolution of \gra{}.
The total error $err$ for $\phi$ and amplitude relative difference $\Delta A/A$
is computed by
$err = (err^2_{\mrm{PIT}}(N^{-p}) + err^2_{\mrm{PIT}}(R^{-p}) + err^2_{\mrm{GRA}}(N^{-p}))^{1/2}$. 
}
\label{fig:err_budget_cce_h_from_psi4}
\end{figure*}
%

\subsection{Numerical methods}
\label{sec:scheme}

We conduct \ac{nr} simulation of \ac{BBH} systems with \gra{}.
This framework computes spatial derivatives using a
sixth-order finite difference scheme, resulting in a truncation error of
$\mathcal O(\delta x^6)$, where $\delta x$ denotes the grid spacing. 
To control numerical noise,
we employ an eighth-order Kreiss-Oliger dissipation
operator with $\epsilon = 0.5$, thereby introducing an error of 
$\mathcal O(\delta x^7)$. 
Time integration is performed using a fourth-order Runge-Kutta
method, incurring an error of $\mathcal O(\delta x^4)$ during extended time
evolutions. Additionally, we maintain a Courant-Friedrichs-Lewy factor of 0.25 
to ensure numerical stability throughout the simulations.

For simulating the \ac{BBH} systems, we use the Z4c formulation of Einstein's
equation~\cite{Bernuzzi:2009ex, Hilditch:2012fp} and set the gauges of the system 
using the moving puncture gauge conditions as specified in \rf{Daszuta:2021ecf}.
\subsection{Grid setup}
\label{sec:grid_setup}

To enhance computational speed through parallelization, we divide the computational
grid~(numerical domain) in \gra{} into multiple subdomains
called \mbs~\cite{Stone2020TheAA,Daszuta:2021ecf}. 

The resolution of the root~(or base) grid is
determined by specifying the number of grid points in each direction.
We divide the root grid into \mbs, each containing a predetermined 
number of grid points that must be a divisor of the total points in the 
root grid. 
Consequently, for a fixed grid extent,
increasing the root grid resolution generates more \mbs{} at the root grid 
thereby increasing the resolution of each \mb.
Then,
these \mbs{} are recursively subdivided to create the 
full adaptive mesh refinement octree structure grid, starting from the root grid.

For adaptive mesh refinement, we employ the $L_2$ refinement method introduced in~\cite{Rashti:2023wfe}, 
in which finer grid levels centre around each \ac{BH}, 
coarsening gradually toward the outer boundary based on an $L_2$-norm criterion. 
As demonstrated in~\cite{Rashti:2023wfe}, this $L_2$ method yields a grid structure 
similar to the \qt{sphere-in-sphere} Berger-Oliger algorithm~\cite{Berger:1984zza}, 
but without overlapping \mbs.

The $L_2$ method provides both speed and accuracy benefits over the originally implemented 
$L_{\infty}$ method~\cite{Daszuta:2021ecf}. By employing the $L_2$ method, 
which generates fewer \mbs{}, we achieve an efficient grid structure as the sphere-like
configurations from the $L_2$ method fit more efficiently within the box structures 
generated by the $L_{\infty}$ method.

In our simulations, the \mb's size,
in all directions, is set to $16$. The number of refinement levels,
set between 12 and 14, is chosen based on the \ac{BBH}'s mass ratio, 
to ensure a minimum resolution 
of $ m_2 /\delta{x} > 25 $ at the lowest resolution of the root grid,
where $m_2$ is the mass of lighter or secondary \ac{BH} and $\delta{x}$ is the grid
spacing at the puncture.
Additionally, the 2:1 refinement constraint further improves resolution in 
the \ac{GW} extraction regions by increasing the refinement level.

We use Sommerfeld boundary conditions~\cite{Hilditch:2012fp} at the outer
boundary of the computational grid. 
These boundary conditions are
imperfect, and may cause reflection of nonphysical modes into the domain.
To mitigate this, we use a sufficiently large domain so that the
largest waveform extraction radius is causally disconnected from the
outer boundary.

Lastly, as all considered binaries exhibit reflection symmetry across
the orbital plane, we only evolve the region above the orbital plane~($z > 0$) 
and impose bitant symmetry at the $z = 0$ boundary.


\subsection{Initial data}
\label{sec:id}

We start our simulations by generating \ac{BBH} puncture initial data using the 
pseudo-spectral elliptic solver \code{TwoPunctures} code~\cite{Ansorg:2004ds}. 
We place the \acp{BH} along the x-axis at a distance $D$ apart, such that the center of mass of each system
with mass ratio $q = m_1/m_2 \geq 1$ is at the origin.
The \code{TwoPunctures} code estimates the component masses $m_1$ and $m_2$
such that the total mass of the binary satisfies $m_1 + m_2 = 1$. 
We use the separation distance $D$ and the momentum components along 
the $x$ and $y$ directions, $P_x$ and $P_y$, from 
\cite{Hannam:2010ec, Ramos-Buades:2018azo}.
This ensures low-eccentricity initial data, producing approximately 
20 quasi-circular pre-merger cycles for the $q=1$ system and around 15 pre-merger 
cycles for $q=2$, $q=3$, and $q=4$ systems.
Table~(\ref{tab:bbh_runs}) summarizes these initial data parameters 
for our simulations.

\subsection{Gravitational waves}
\label{sec:grav_wave}
To compute waveforms at null infinity $\mathcal{I}_+$, we use either an approximate
method based on post-Newtonian theory (\acl{FRE}; FRE)\acused{FRE}, or
the \ac{CCE} method. The latter directly solves Einstein's equations to $\mathcal{I}_+$,
and therefore represents our preferred technique for computing \acp{GW}. We
describe it in detail in this section.
Additional information regarding the \ac{FRE} method can be found in \app{sec:fre_method}, 
while \app{sec:fre_vs_cce} collects comparisons of data obtained with the
two methods to evaluate their systematic uncertainties.

To generate \acp{gw} at $\mathcal{I}_+$ using the publicly available \pit{} code~\cite{Bishop:1998uk, Babiuc:2010ze}, 
we develop a pipeline that generates the necessary input data. 
This pipeline expands the metric fields—specifically the spatial metric \(\gamma_{ij}\), the 
shift vector \(\vec{\beta}\), and the lapse \(\alpha\)—in the \(3+1\) formalism~\cite{gourgoulhon_thebook} 
over a defined radial range \([r_1, r_2]\). Each metric field $\mathcal{G}$ is expanded as follows:
\be
\mathcal{G}(t, x^i) = 
\sum^{k_{\mrm{max}}}_{k=0}
\sum^{l_{\mrm{max}}}_{l=0}
\sum^{l}_{m=-l} 
C_{klm}(t) U_k(\tau(r)) Y_{lm}(\theta, \phi),
\label{eq:cce_metric_exp}
\ee
where, $U_k(\tau)$ are Chebyshev polynomials of the second kind,
$Y_{lm}(\theta,\phi)$ are spherical harmonic basis functions, and
\be
\tau(r) := \frac{2r - r_1 - r_2}{r_2 - r_1}.
\ee

For this study, we use \(k_{\mathrm{max}} = l_{\mathrm{max}} = 11\). 
The expansion coefficients \(C_{klm}\) are computed every $20^\mathrm{th}$ to $30^\mathrm{th}$ time
steps during the evolution and stored in an \texttt{hdf5} file to serve as boundary condition samples
on concentric world tubes for the \pit{} code. 

\app{sec:cce_fourier_filter} shows, as an example, the $C_{222}$ coefficients for $\gamma_{xx}$.
The presence of noise in the $C_{klm}$ coefficients 
at the beginning of the run and even at a later times 
is one of the challenges to deal with when computing \ac{CCE}
strains. We have tested different filtering methods to address this issue,
as detailed in \app{sec:cce_fourier_filter}, but ultimately chose not to 
filter $C_{klm}$ when computing the presented waveforms. 

We can use the Weyl curvature scalar $\mpsi$ or the News function $\mnews$ to compute strains as
the \pit{} code computes both. These are related to the \ac{gw} strain as:
\begin{equation}
  \mpsi = \ddot{h}_+^{\ell m} - i\ddot{h}_\times^{\ell m},\quad
    \mnews = \dot{h}_+^{\ell m} - i\dot{h}_\times^{\ell m},
\label{eq:psi4_news}
\end{equation}
where the dots indicate time derivatives. In both scenarios, the initial step involves normalizing the
quantities by multiplying them by a factor of $2\times(-1)^{l+m}$.
To recover the strain from either, we use the \ac{FFI} method~\cite{Reisswig:2010di}. 
The cutoff frequency employed for this process is given in
\app{sec:ffi_method}.

It is important to note that while our \ac{CCE} strains include \ac{GW} 
memory effects~\cite{Blanchet:1992br}, the evaluation of these effects 
requires the complete past history of the spacetime. 
During the \ac{FFI} process, we lose the integration constant that represents 
the history of the mode prior to the start of the \ac{nr} simulation. 
Consequently, the \ac{FFI} method does not account for 
this constant of integration related to the memory effect.
Additionally, our choice of cutoff frequency is optimized for 
the oscillatory modes of \ac{GW} strains. 
Readers may refer to~\cite{Pollney:2010hs}, where \ac{GW} strains are 
matched with post-Newtonian strains to incorporate \ac{GW} memory effects. 
Furthermore, we do not apply any supertranslations to 
transform the strains into the appropriate BMS frame, 
as discussed in~\cite{Mitman:2024uss}.

\subsection{Uncertainty quantification}
\label{sec:conv_test}

We estimate the finite-resolution and finite-extraction radius errors in our \ac{nr} simulations 
by performing a series of convergence studies on the strain data obtained using \ac{CCE}, as discussed in 
Sec.~\ref{sec:cce_convergence}, and \ac{FRE}, as detailed in \app{sec:fre_convergence}.
In both cases, we model the error in any numerically computed quantity $w$ as a 
power of $\delta$~\cite{Richardson:1911, alcubierre_book} 
\be
  w(\delta) = w_{\rm ex} + c_n \delta^n + \mathcal{O}\big(\delta^{n+1}\big).
  \label{eq:q_expand}
\ee

Here, $w_{\rm ex}$ represents the exact value of $w$. 
The parameter $\delta$ can be $N^{-1}$, 
where $N$ is the number of grid points on the root grid 
of the simulation when evaluating finite-resolution errors;
$\delta$ can also be $R^{-1}$, where $R$ is the extraction radius, 
when assessing finite extraction-radius effects.
The coefficient $c_n$ quantifies the amplitude of the leading-order error term.
Note that in Equation \eq{eq:q_expand}, no summation is implied over the same indices.

To estimate the convergence order, we exploit the difference in $w$ obtained at 
various $\delta$'s
\be
\big| w(\delta_{1}) - w(\delta_{2}) \big| - 
\mathcal{C}(n) \big| w(\delta_{2}) - w(\delta_{3}) \big|,
 \label{eq:conv_order}
\ee
where $\mathcal{C}(n)$ is defined
\be
  \mathcal{C}(n) =
  \left|\frac{ \delta_{1}^{n} - \delta_{2}^{n} }{ \delta_{2}^{n} - \delta_{3}^{n}
}\right|.
 \label{eq:conv_order_c}
\ee
By minimizing \eq{eq:conv_order} with respect to $n$, we infer the value of $n$ that characterizes
the optimal convergence rate.

Moreover, for a specific $\delta$, for instance, $w(\delta_2)$, 
we can estimate the magnitude of the leading order error, i.e., $|c_n \delta^n|$,
using
\be
    \left|\frac{w(\delta_{1}) - w(\delta_{2})}
{1-\left(\frac{\delta_{1}}{\delta_{2}}\right)^{n}}\right|.
    \label{eq:trunc_error}
\ee
If no convergence is observed for a specific study, we use $n=0$; in these cases, 
the error is estimated by $\left|w(\delta_{1}) - w(\delta_{2})\right|$.

For example, when evaluating finite-resolution errors pertinent to the
\gra{} resolution, $\delta = N^{-1}$ and
$N$ can take values such as 128, 192, 256, 320, and 
384~(see Table~\ref{tab:bbh_runs}). Once we determine $n$, by minimizing
\eq{eq:conv_order}, and $c_n$, by using \eq{eq:trunc_error}
we can quantify the leading-order error term, 
namely $c_n \delta^n$, for each given resolution. 
Similarly, when assessing finite extraction-radius effects, 
$R$ can assume values like 20, 50, and 100, 
allowing us to estimate the error associated with the radius of the world tube in the waveforms.

Lastly, in our convergence studies, after the initial junk radiation,
i.e., Brill waves~\cite{alcubierre_book,Brill:1959zz},
we observe some dephasing between different resolutions of the same run.
Therefore, we align the waveforms in terms of phase and time at the maximum
amplitude peak.
This ensures that all waveforms reach their maximum amplitude 
simultaneously and share the same phase value.
This alignment is essential for conducting convergence analysis 
across different extraction radii, as the timing of the maximum peak varies.

\section{Results}
\label{sec:results}

\subsection{Simulations}
\label{sec:simulations}
After constructing the initial data of our \ac{BBH} systems using
the \code{TwoPunctures} code, as explained in \ref{sec:id}, 
we dynamically evolve the \ac{BBH} systems using the \gra{} code.

The total simulation times chosen for each mass ratio are as follows: 
$q=1$ with the total time $\sim 2370$,
$q=2$ with the total time $\sim 1370$,
$q=3$ with the total time $\sim 1566$, and
$q=4$ with the total time $\sim 1745$.
The computational cost of these 19 simulations, table~(\ref{tab:bbh_runs}), 
was approximately $155$ million core-hours,
with a code speed of about 
$8\times10^7$ zone-cycle updates per second. 
This includes significant file system operations required to store 
the $C_{klm}$ coefficients of \eq{eq:cce_metric_exp}, 
which are necessary for computing the \ac{CCE} strains.

The trajectories for each puncture, $\vec{x}_{p}(t)$, 
are found by solving the ordinary differential equation
$\dot{\vec{x}}_{p}(t) = -\vec{\beta}_p(t)$,
in which, $\vec{\beta}_p(t)$ denotes the shift vector at the puncture~\cite{Daszuta:2021ecf,Campanelli:2005dd}.
The initial values of $\vec{x}_{p}(t=0)$ are used from the
initial data parameters, and hence by solving this ordinary differential
equation, we can find the trajectory of each \ac{BH} during the evolution.
This method is a standard method and also used in e.g.,
\texttt{Einstein Toolkit}~\cite{Campanelli:2005dd}.
Additionally, in \texttt{AthenaK}~\cite{Zhu:2024utz}, we have experimented with 
tracking the minimum of the lapse on the grid, but it is less accurate
than solving the the ordinary differential equation for $\vec{x}_{p}(t)$.

We estimate the eccentricity of our simulations shown in Fig.~\ref{fig:orbits} using the methods discussed in \citep{Tichy:2019ouu}.
This approach requires us to identify parameters, including eccentricity, 
to fit the coordinate distance between punctures, 
derived from the trajectories $\vec{x}_{p}(t)$, as a function of evolution time.
We use the initial orbital angular velocity and the first few orbits of the
simulation to find a fit for the coordinate distance versus time.
We choose the coordinate distance and not its time derivative, as
done in~\cite{Pfeiffer:2007yz}, because the time derivative of the coordinate distance can
introduce noise.
In particular, we use $\sim 4$ orbits and calculate the initial orbital velocity using the
initial momenta and distance between the puncture, see \app{sec:ffi_method}.
Specifically, \ac{BBH} runs with $q=$ 2, 3, 4 have an 
eccentricity of $\sim 10^{-3}$, and the $q=1$ case has an eccentricity of $\sim 6\times10^{-4}$. 

\fig{fig:h_catalog} shows the $(\ell=2,m=2)$ and $(\ell=4,m=4)$ modes of the 
\ac{CCE} strains for our simulations. To ensure that the merger happens at $t=0$, 
we subtract the merger time $t_{\mrm {mrg}}$, defined as the peak amplitude time of the $(2,2)$ mode, in the time axis.
%
At $t_{\mrm {mrg}}$, a highly excited \ac{BH} is formed, which then settles down 
to its stationary Kerr state by emitting \acp{gw}.
In \fig{fig:h_catalog}, we see each system exhibits a clear inspiral phase, 
followed by a merger and a ringdown. The subdominant $(4,4)$ mode also captures these 
key features, albeit with a smaller amplitude.

\subsection{CCE error analysis}
\label{sec:cce_convergence}

After computing the \ac{CCE} strains from $\mpsi$, 
using \ac{FFI}~(\app{sec:ffi_method}), we want
to determine the convergence order, and hence finding an estimate of
the error budget in the waveforms.

As described earlier in Sec.~(\ref{sec:grav_wave}), we save $C_{klm}(t)$
coefficients at different radii during \gra{} evolution run, and then we
deploy the \pit{} code using the boundary condition samplings provided by 
the $C_{klm}(t)$ coefficients to solve the \ac{CCE} equations. As such,
we have 3 sources that contribute to the error budget:
the resolution of \pit{} runs,
the world tube radius used to compute $C_{klm}(t)$ coefficients, and
the resolution of the evolution run in 
\gra{}~(which affects the boundary condition for \pit{}).
We study the effect of these sources separately, and find the 
total error in \ac{CCE} waveforms by summing the errors in quadrature.

To find the errors due to the \pit{} resolution and the \pit{} world tube radius, 
we use the $C_{klm}(t)$ data belonging to the highest available resolution
of \gra{} for each mass ratio, and with various extraction tube radii, i.e., 
$20$, $50$, and $100$.
Then, for each radius, we run \pit{} for three different resolutions with a
time step corresponding to that resolution, as outlined
in~\cite{Babiuc:2010ze}.
Specifically, we use
the resolution $224\times 200\times 200$ with the time step $\Delta t$,
the resolution $112\times 100\times 100$ with the time step $\Delta t/2$, and
the resolution $56\times 50\times 50$ with the time step $\Delta t/4$.
Here, the resolution is denoted by 
$N_r \times N_p \times N_q$, in which $r$ is the radial direction
and $(p,q)$ are the stereographic coordinates in \pit{}.
$\Delta t$ is the writing time of $C_{klm}(t)$, i.e.,
20-th to 30-th of \gra{} evolution time, mentioned in Sec.~(\ref{sec:grav_wave}).

We first find the convergence order with respect to the \pit{} resolution.
Accordingly, we use \eq{eq:conv_order} with $\delta=N^{-1}$, where $N$ refers to the $N_r$ 
resolution of \pit{} runs.
\fig{fig:conv_test_cce_h_from_psi4} depicts the convergence of \ac{CCE}
strain's phase computed from $\mpsi$ with a world tube radius of $50$~%
\footnote{Among different world tube extraction radii, only radius = $50$
exhibits convergence behavior with respect to \pit{} resolutions.}.
We observe a 2nd order convergence with respect to the \pit{}
resolution. Similary, we conduct a similar convergence study for \ac{CCE}
strain's amplitude and find a 2nd order convergence in all runs except $q=1$
that exhibits a 1st order convergence.

For the convergence order with respect to the \pit{} extraction world
tube, we use the highest resolution of \gra{} and \pit{}, 
and use \eq{eq:conv_order} with $\delta=R^{-1}$, in which $R$ refers to
the world tube radius extraction of $20$, $50$, and $100$. 
The result of this convergence study,
depicted in \fig{fig:conv_test_rex_cce_h_from_psi4},
shows that the \ac{CCE} strain's phase
demonstrates a consistent $\sim$ 2nd order convergence for all mass ratios.
However, the amplitude converges in different orders for different mass
ratios. In particular, the amplitude in $q=1$ shows no convergence, $q=2$
converges with a 2nd order rate, and $q=3$ and 4 converge with a 3rd order
rate.

Assessing the error budget of the waveforms due to \gra{}, we first need
to find the convergence order with respect to \gra{} resolutions.
As such, we run \pit{} for three $C_{klm}(t)$ data which correspond to 
three resolutions from table~(\ref{tab:bbh_runs}). 
Specifically, we choose \gra{} resolutions of 
384, 256, 128 for $q=$ 1, 2, 3, and 
320, 256, 128 for $q=$ 4, to see a clear convergence pattern.
Then, we run \pit{} with the resolution $224\times 200\times 200$ and the time step $\Delta t$ 
with the world tube radius of $50$ for each given \gra{} resolution.
Lastly, we use \eq{eq:conv_order} with $\delta=N^{-1}$ to find the convergence
order, where $N$ denotes the selected \gra{} resolutions.
We find that $q=1$ and $q=2$ runs, respectively, show a 4th order convergence and
a 2nd order convergence in the phase, while the higher mass ratio runs of $q=3$ and
$q=4$ do not converge with respect to the \gra{} resolutions. Similarly, for
mass ratio runs of $q=$ 1 and 2, the amplitude shows a 4th order
convergence, while the higher mass ratios do not converge.

Having found all convergence orders, now we can use \eq{eq:trunc_error} 
to estimate the errors associated with each source, namely, 
the \pit{} resolution, the extraction world tube radius, and the \gra{} resolution.
As such, we subtract the phases and the amplitudes of two different \ac{CCE}
strains~(different in resolution or extraction radius) in the time domain
to compute the numerator of \eq{eq:trunc_error}.
To carry out this subtraction, we first align the strains and then calculate 
their phase and amplitude differences. The alignment is done by finding a
global time and phase shift $(\Delta\phi_{22}, \Delta t)$ such that
\begin{equation}
  \label{eq:phasing}
  \chi^2 = \int_{t_i}^{t_f} 
    [\phi_{h_1}(t + \Delta t) - \phi_{h_2}(t) - \Delta\phi_{22}]^2 dt \, ,
\end{equation}
is minimized.
In \eq{eq:phasing}, $\phi_{h_1}, \phi_{h_2}$ are the $(\ell, m) = (2,2)$ \ac{gw} phases of 
the two waveforms $(h_1,h_2)$ to be compared and $(t_i, t_f)$ specify 
the time window over which the minimization is
performed. We place the time window after the junk radiation. 

We summarize the results of the convergence study as well as the error
analyses,  for both amplitude and phase of the (2,2) 
mode, in \fig{fig:err_budget_cce_h_from_psi4}.
Specifically, using \eq{eq:trunc_error}, 
we estimate the magnitude of the leading order error term, 
expanded in \pit's resolutions, 
at the highest resolution of \pit{} run; 
we denote this error as  $err_{\mrm{PIT}}(N^{-p})$, 
where $p$ shows the convergence order.
Similarly, we denote the error due to radius extraction at
$R=50$ as $err_{\mrm{PIT}}(R^{-p})$,
and, the error due to \gra's resolutions, 
at the highest resolution run of \gra, as $err_{\mrm{GRA}}(N^{-p})$.
Additionally, we represent the total error with a solid blue line. 
This total error is calculated as the quadrature sum of the individual
errors.

In \fig{fig:err_budget_cce_h_from_psi4}, we observe that the \pit{} resolution
error for the phase, denoted as $err_{\mrm{PIT}}(N^{-p})$,
is the smallest error among various sources of error for
all mass ratios except $q=1$. For amplitude, $err_{\mrm{PIT}}(N^{-p})$ is
smallest only for mass ratios $q=3$ and $q=4$. 
The phase error shows a slight increase as we approach the merger,
ranging from $(10^{-5},\sim 10^{-3})$ for different mass ratios.
In contrast, the error in amplitude remains relatively constant 
for $q=$ 2, 3, and 4 cases, and it is slightly increasing only for $q=1$. 
The range of this error is in $(10^{-4},\sim 5\times 10^{-2})$.
The observed convergence, with respect to the \pit{}'s resolution, suggests
$err_{\mrm{PIT}}(N^{-p})$ decreases as the resolution increases.
For instance, \rf{Barkett:2019uae} finds a maximum error of $\sim 10^{-6}$ in a
bouncing \ac{BH} system when the resolution of $900\times 900 \times 900$ was
used in \pit.

The error of the extraction radius is denoted by $err_{\mrm{PIT}}(R^{-p})$
in \fig{fig:err_budget_cce_h_from_psi4}.
We observe that the largest source of error is due to the extraction radius for
both phase and amplitude.
To mitigate this error, one can use a larger extraction radius. 
For example, by using a radius of $100$, the error term reduces by a factor of
4, cf. \eq{eq:trunc_error}. However, only the strains with the extraction
radius of 50 show a consistent convergence order.
Moreover, as we discuss in Sec.~(\ref{sec:mismatch}), the strains with
the extraction radius of $R=50$ exhibit much smaller mismatch than $R=100$.
Consequently, we select to use strains with world tube execration $R=50$ for
our analysis.
It is worth noting that while theoretically the \ac{CCE} strains are solutions 
to the full Einstein equations, and thus should not depend on the extraction radius, 
practical considerations complicate this expectation. 
Specifically, in the \pit{} code, there is an assumption~\cite{Babiuc:2010ze} 
regarding the geometry of the initial null hypersurface, which is presumed 
to be close to a Schwarzschild geometry. 
This assumption is necessary for setting the initial characteristic data, 
making it less accurate for smaller extraction radii.
Conversely, larger radii can introduce challenges due to noise present 
in our data, as discussed in \app{sec:cce_fourier_filter}, 
as well as numerical errors arising from small corrections to a flat geometry
which may not be adequately captured by our Cartesian grid.

In \fig{fig:err_budget_cce_h_from_psi4}, we see that the error in the phase and
amplitude associated to the \gra{} resolution, $err_{\mrm{GRA}}(N^{-p})$,
is relatively larger during the inspiral for high mass ratio runs compared
to lower mass ratio runs with $q=$ 1 and 2.
Moreover, this error grows as it gets close to the merger,
and it is within the range of $(10^{-5},10^{-2})$ for both phase and
amplitude. 

For $q=$ 2, 3, and 4, we observe in \fig{fig:err_budget_cce_h_from_psi4} 
that during the inspiral, 
and after junk radiation, the total error for 
the dephasing is less than 0.01 radians. Moreover,
the total error for the amplitude relative difference
is below $\sim 1\%$. 
As we get close to the merger, the total error for these cases 
increases but still remains below 0.05 radians for the phase and below $\sim 5\%$
for the amplitude relative difference.
For $q=1$, we see a larger total error, 
in its phase and amplitude, with respect to the higher mass ratio runs.

Lastly, we find that the \ac{CCE} strains derived from $N^\lm$ show no discernible
convergence pattern for neither the resolution nor the world tube extraction
radius. Consequently, we have opted not to include these results in our analysis.

\subsection{Self-mismatch}
\label{sec:mismatch}
%

Presuming our highest resolution run produces the most accurate waveform, 
we assess the accuracy of our \ac{nr} simulations for different extraction
radii by comparing the agreement between this waveform and the waveforms 
generated from simulations with identical initial data but lower resolutions.
To quantify this, we use the noise-weighted 
inner product between two Fourier-domain waveforms, \( h \) and \( h' \):
\begin{equation}
(h \mid h') = 
    4\,\mathrm{Re}\,\int^{f_\mathrm{high}}_{f_\mathrm{low}} 
        df \, \frac{\tilde{h}^\ast \tilde{h'}}{S_n(f)},
\end{equation}
for a detector noise power spectrum, $S_n(f)$ and compute the overlap \(\mathcal{O}\) 
between the two waveforms~\cite{Harry:2017weg, Chandra:2022ixv}:
%
\begin{align}
\label{eq:skymax-overlap}
& \mathcal{O}(h,h') = \\ \nonumber
& \max_{\Delta t_c, \Delta u} \frac{( \hat{h}' | \hat{h}_+)^2 + 
    ( \hat{h}' | \hat{h}_\times)^2 - 2( \hat{h}' | \hat{h}_+)
    ( \hat{h}' | \hat{h}_\times)(\hat{h}_+| \hat{h}_\times)}
    {1-(\hat{h}_+ | \hat{h}_\times)^2}.
\end{align}
%
Here, \( \hat{h}_{+/\times} = \frac{h_{+/\times}}{\sqrt{(h_{+/\times} \mid h_{+/\times})}} \) 
represents the unit-normalized polarization, and
\[
u = \frac{F_{+} \sqrt{\left(h_{+} \mid h_{+}\right)}}{F_{\times} \sqrt{\left(h_{\times} \mid h_{\times}\right)}}
\]
depends on the detector response function \( F_{+/\times} \). The maximisation over the peak time difference $\Delta t_c$ is performed
efficiently using an inverse fast Fourier transform routine. Thus, by definition, for a chosen inclination angle $\iota$ \eq{eq:skymax-overlap} is effectively maximized over an overall amplitude scaling, time differences, sky location, and polarization angle. 
The overlap \(\mathcal{O}=1\) if the two waveforms are identical up to a re-scaling; otherwise, \(\mathcal{O} < 1\). 
This reduces to:
\begin{equation}
    \mathcal{O}(h,h') = \max_{\Delta t_c} (\hat{h}' \mid \hat{h}_+)
\end{equation}
when \(\hat{h}_+ = i\hat{h}_\times\), as is the case when comparing the individual modes \((\ell,m)\) for each waveform.

It must be noted that \eq{eq:skymax-overlap} does not account for the two waveforms' phase differences, $\Delta \phi$ that can affect the distinguishability or ``mismatch'' between the two waveforms.
Therefore, for each pair of waveforms, we numerically maximize over $\Delta \phi$ and calculate the mismatch:
\begin{equation}\label{eq:mismatch}
    \mathcal{M}\left(h, h'\right)=1-\max _{\Delta \phi} \mathcal{O}(h, h')
\end{equation}
which measures the inconsistency between the two \ac{nr} waveforms. 
Consistent with the rest of the paper, we use the strains
obtained using the \ac{CCE} method with the world tube extraction radii of 50
and 100.

Our goal is to assess the mismatch of the highest resolution run in
table~(\ref{tab:bbh_runs}) against the `exact' benchmark, i.e., an infinite
resolution run, for
the next-generation detectors such as LISA~\cite{LISA:2017pwj},
the \ac{et}~\cite{Maggiore:2019uih}, and
\ac{ce}~\cite{Reitze:2019iox}. 
Whenever we evaluate the mismatches for these detectors, we employ the frequency intervals
given by the starting frequency of the \gra{} waveform $f_{\ell m}^{0} = m/2 \omega_{22}/(2\pi)$ 
(multiplied by 1.25, to avoid artifacts due to FFTs) and $1024$ or $1$ Hz, respectively for \ac{ce} and LISA.
Additionally, when higher modes are considered, we fix the inclination angle to $\iota = \pi/3$.
%
%
%
To this aim, we first model the mismatch between lower resolutions with respect to 
the highest available resolution in table~(\ref{tab:bbh_runs}) using
\be
\mathcal{M}\left(p,c,\delta x,\delta x_{\mrm ref}\right) = 
c \left ({\delta x}^p-\delta x_{\mrm ref}^{p}\right),
\label{eq:mismatch_model}
\ee
where $p$ and $c$ are coefficients to be fit using the data, 
and $\delta x_{\mrm ref}$ is the resolution of the benchmark run, 
cf.~\cite{Vaishnav:2007nm,Ferguson:2020xnm},

Using this ansatz, we can then estimate the mismatch between a waveform at a given 
resolution $\delta x$ and the exact solution as 
$\mathcal{M}(c,p,\delta x,\delta x_{\mrm ref}=0)$.

\begin{figure*}[t]
\centering
\includegraphics[width=0.48\linewidth,clip=true]{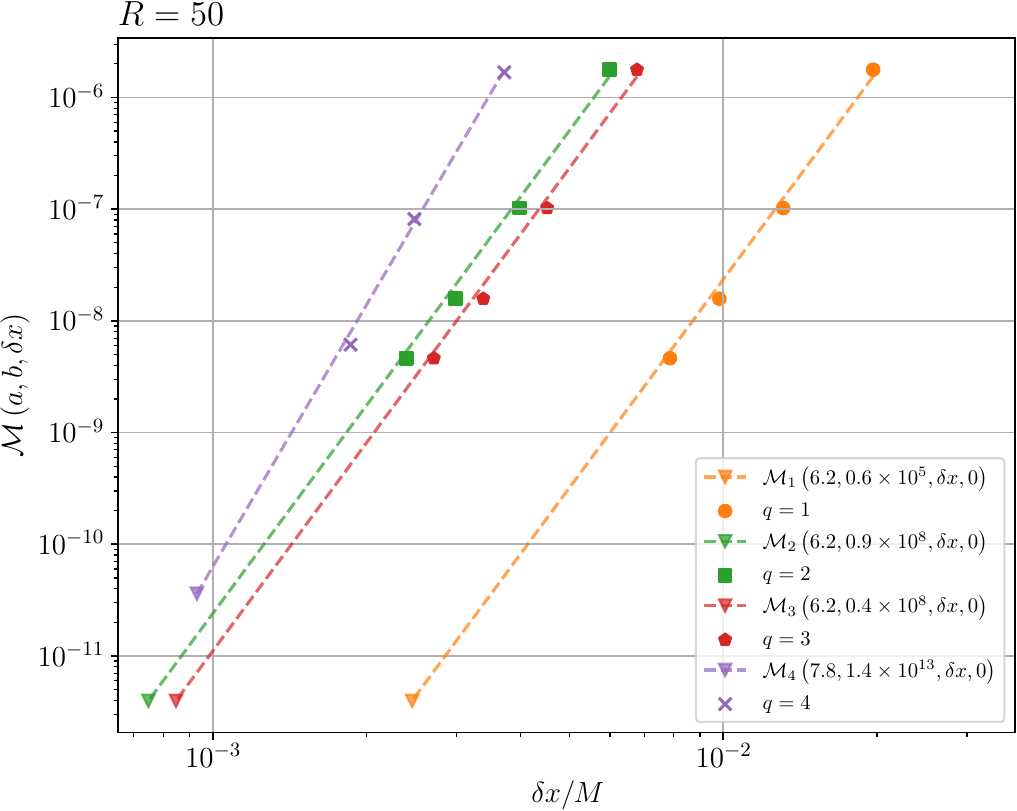}
\includegraphics[width=0.48\linewidth,clip=true]{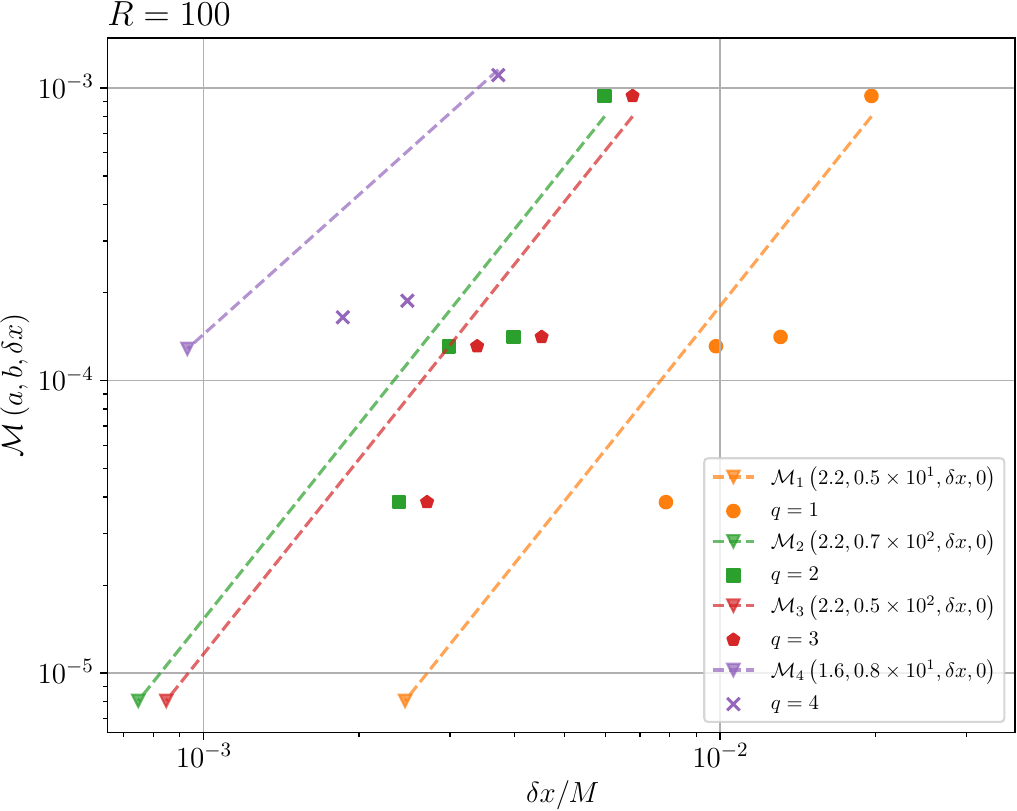}
\caption{%
The discrete points for each mass ratio show the mismatch between 
low resolution and the highest resolution of \ac{CCE} strains with $(2,2)$ mode.
The dashed lines estimate the limit of the \eq{eq:mismatch_model} at 
$\delta x_{\mrm ref} \to 0$. The upside-down triangle at the end of each dashed
line specifies the mismatch between the highest resolution and the exact
resolution.
In the left panel we use $C_{klm}(t)$ coefficients extracted at $R=50$ to compute \ac{CCE}
strains, and in the right panel, the coefficients extracted at $R=100$.
}
\label{fig:mismatch_strain_cce}
\end{figure*}

\fig{fig:mismatch_strain_cce} shows the mismatch of the \((2,2)\) mode strains obtained using the \ac{CCE} method
for two different world tube extraction radii of 50, the left panel, and 100,
the right panel.
We employ the LISA noise curve~\cite{Babak:2021mhe} for this comparison and consider a total mass of \(10^5 M_\odot\). 
The discrete data points represent the mismatch between the highest and the other resolutions.
Additionally, the corresponding fit, described by \eq{eq:mismatch_model} with 
$\delta x_{\mathrm{ref}} \to 0$, 
is shown as a dashed line, with different colors indicating each mass ratio. 
An upside-down triangle on the dashed line indicates the mismatch between the highest 
resolution and the exact solution, 
i.e., evaluating \eq{eq:mismatch_model} with the fitted values of $c$ and $p$ 
and setting $\delta x_{\mrm ref}=0$.

It is important to note that LISA can detect signals with SNR as high as 1000, 
which requires a mismatch of \(\lesssim 10^{-7}\). This is because the
extremised loudness difference, 
$(\delta h \mid \delta h)  = (h - h' \mid h - h')$, between the waveforms $h'$ and $h$, 
depends on the signal's SNR as \(2\rho^2(1-\mathcal{M}(h,h'))\). At an SNR of 1000, 
this difference, $(\delta h \mid \delta h) \sim 0.2$, remains below the whitened noise variance of 1, 
indicating any difference between the waveforms will not manifest as a difference in binary parameter 
measurement. 

We see in the left panel of \fig{fig:mismatch_strain_cce} 
that the mismatch between the highest resolution run and the second
highest resolution run at $R=50$ is $\sim 5 \times 10^{-9}$ for all mass ratios.
Additionally, we observe that for mass ratios \(q=1\), \(q=2\), and \(q=3\), 
the mismatch of our highest resolution strain against the \qt{exact} strain 
is $\sim 10^{-12}$,
while the \(q=4\) run exhibits a mismatch $\sim 10^{-11}$.
The latter mismatches are orders of magnitude more accurate
than the $\sim 10^{-7}$ threshold required by LISA.

In the right panel of \fig{fig:mismatch_strain_cce}, we present the mismatch
between the waveforms extracted at $R=100$.
The mismatch between the highest resolution strain and the second
highest resolution strain is $\sim 3 \times 10^{-5}$ for $q=$ 1, 2, and 3
runs, and $\sim 10^{-4}$ for $q=4$. 
The mismatch of the highest resolution strain against the \qt{exact} strain, 
namely, when $\delta x_{\mrm ref} \to 0$ in \eq{eq:mismatch_model}, 
is $\sim 10^{-5}$ for mass ratios of 1, 2, and 3, 
while it is $\sim 10^{-4}$ for the $q=4$ run.
These mismatches are significantly
larger than those observed in the left panel and do not meet 
the mismatch requirement set by LISA.

Lastly, our mismatch results quantitatively remain unchanged for total masses 
in the range \((10^4, 10^6) M_\odot\).

%

\subsection{Comparison with SXS waveforms}

\begin{figure*}[t]
  \includegraphics[width=\linewidth]{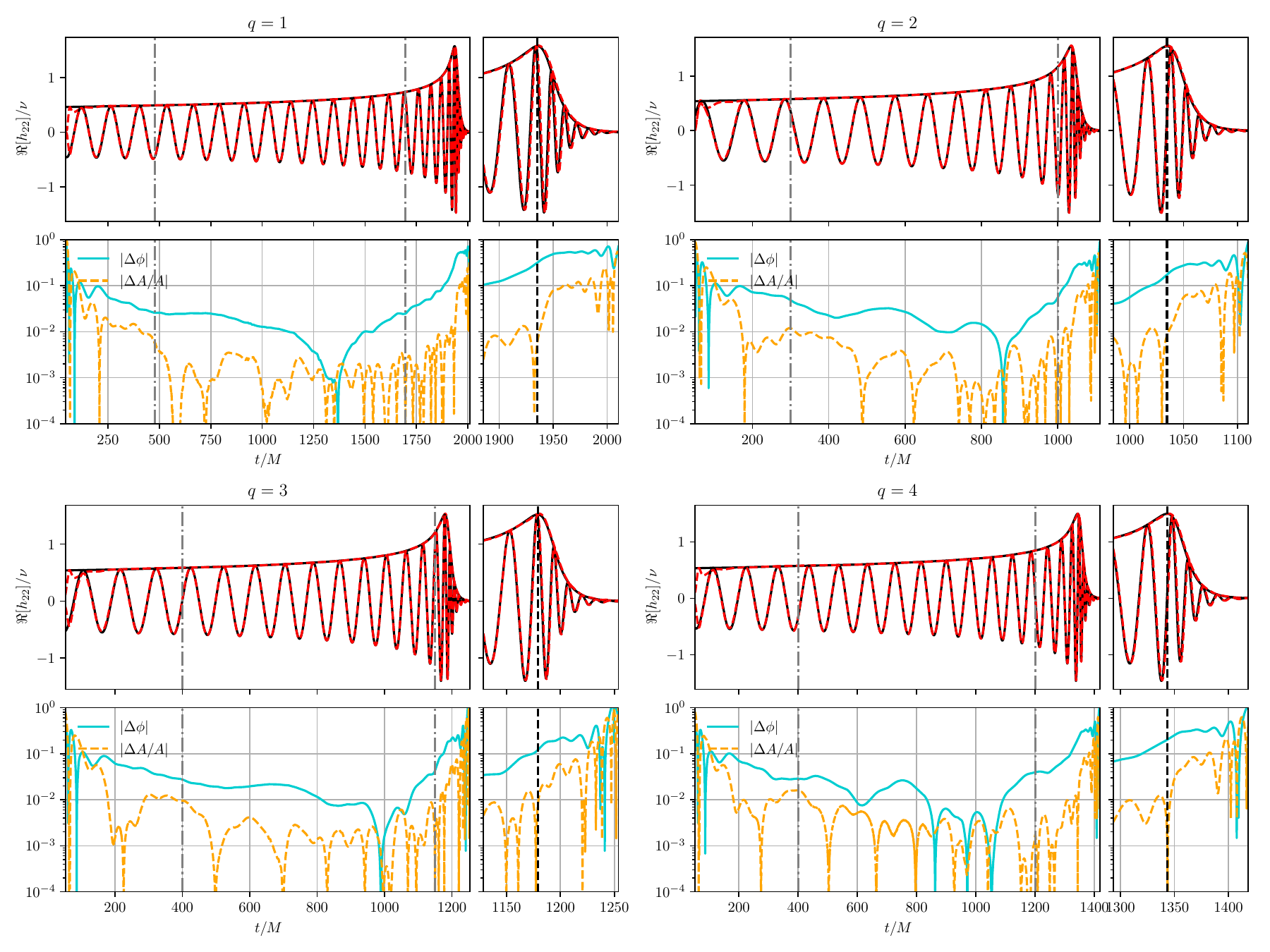}
  \caption{Comparison of the $h_{22}$ mode between the \gra{} (red dashed) and SXS (black) simulations for all four mass ratios considered. 
  Top panels show the real part of the waveform multipoles, while the bottom panels display the phase difference 
  $\Delta\phi = \phi_{\ell m}^{\rm SXS} - \phi_{\ell m}^{\rm GRA}$ and the amplitude relative difference 
  $\Delta A / A= 1 - A_{\ell m}^{\rm GRA} / A_{\ell m}^{\rm SXS}$. The vertical dashed-doted lines indicate the interval
  used for the alignment, while the vertical dashed lines mark the merger time. We find that phase differences at merger and during the 
  inspiral rarely exceed $\sim 0.1$ rad, with the $q=4$ case being the only exception, and amplitude relative differences typically below $1\%$.}
  \label{fig:sxs_22_allq}
\end{figure*}

\begin{figure*}[t]
  \includegraphics[width=\linewidth]{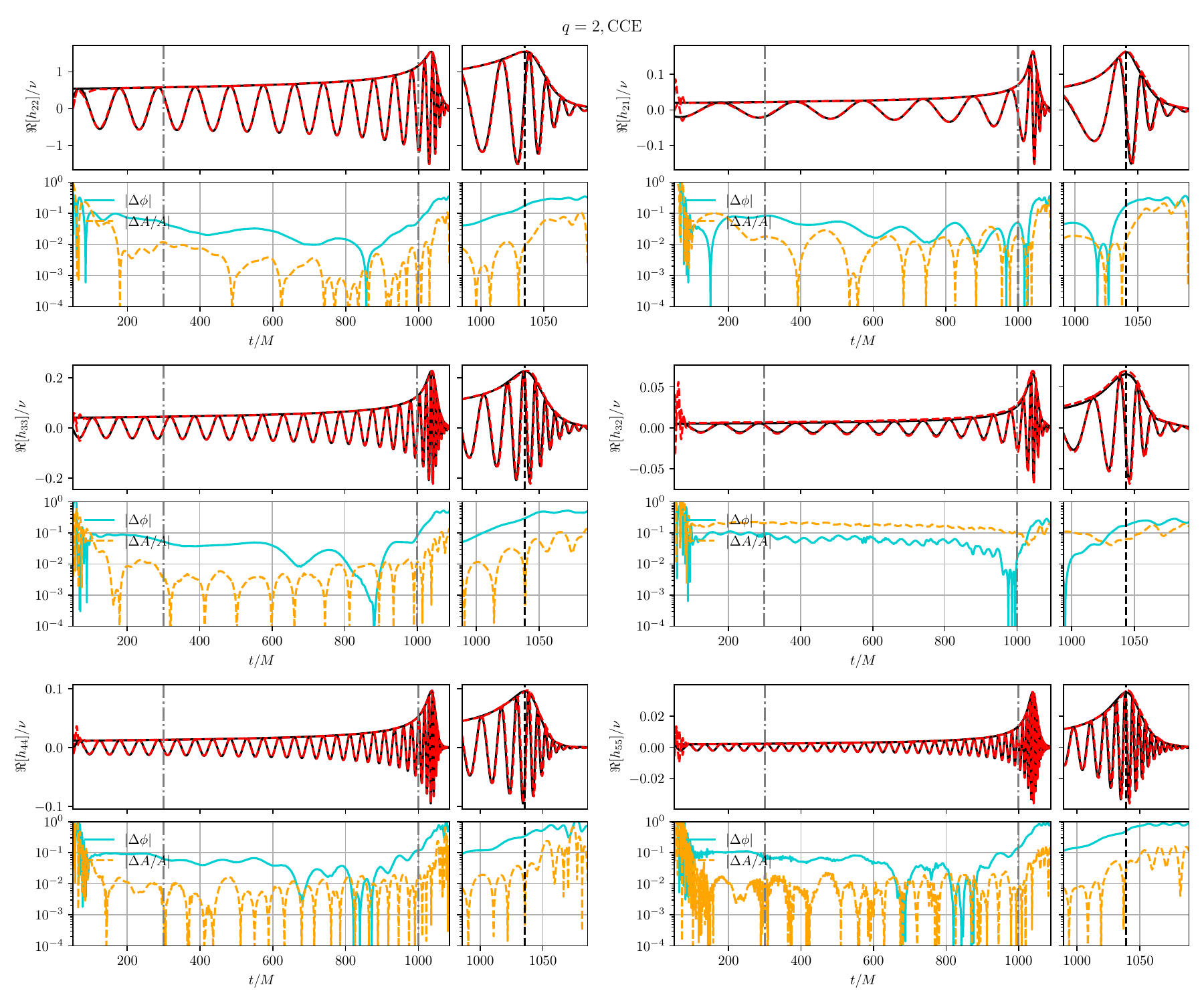}
  \caption{Comparison of \gra{} (red dashed) and SXS (black) waveforms for the $(\ell, m) = \{(2,2), (2,1), (3,3), (3,2), (4,4), (5,5)\}$ 
  modes and $q=2$. Top panels show the real part of the waveform multipoles, while the bottom panels display the phase difference $\Delta\phi = \phi_{\ell m}^{\rm SXS} - \phi_{\ell m}^{\rm GRA}$
  and the amplitude relative difference $\Delta A = 1 - A_{\ell m}^{\rm GRA} / A_{\ell m}^{\rm SXS}$. The vertical dashed-dotted lines indicate the interval
  used for the alignment of the $(2,2)$ mode, while the vertical dashed lines mark the merger time. We find overall excellent agreement between the two
  waveforms, with phase differences at merger and during the inspiral rarely exceeding $\sim 0.1$ rad. 
  The largest differences are observed for the $(3,2)$ merger-ringdown portion. See the text for further details.}
  \label{fig:sxs_td_q2}
\end{figure*}

\begin{figure*}[t]
  \includegraphics[width=\linewidth]{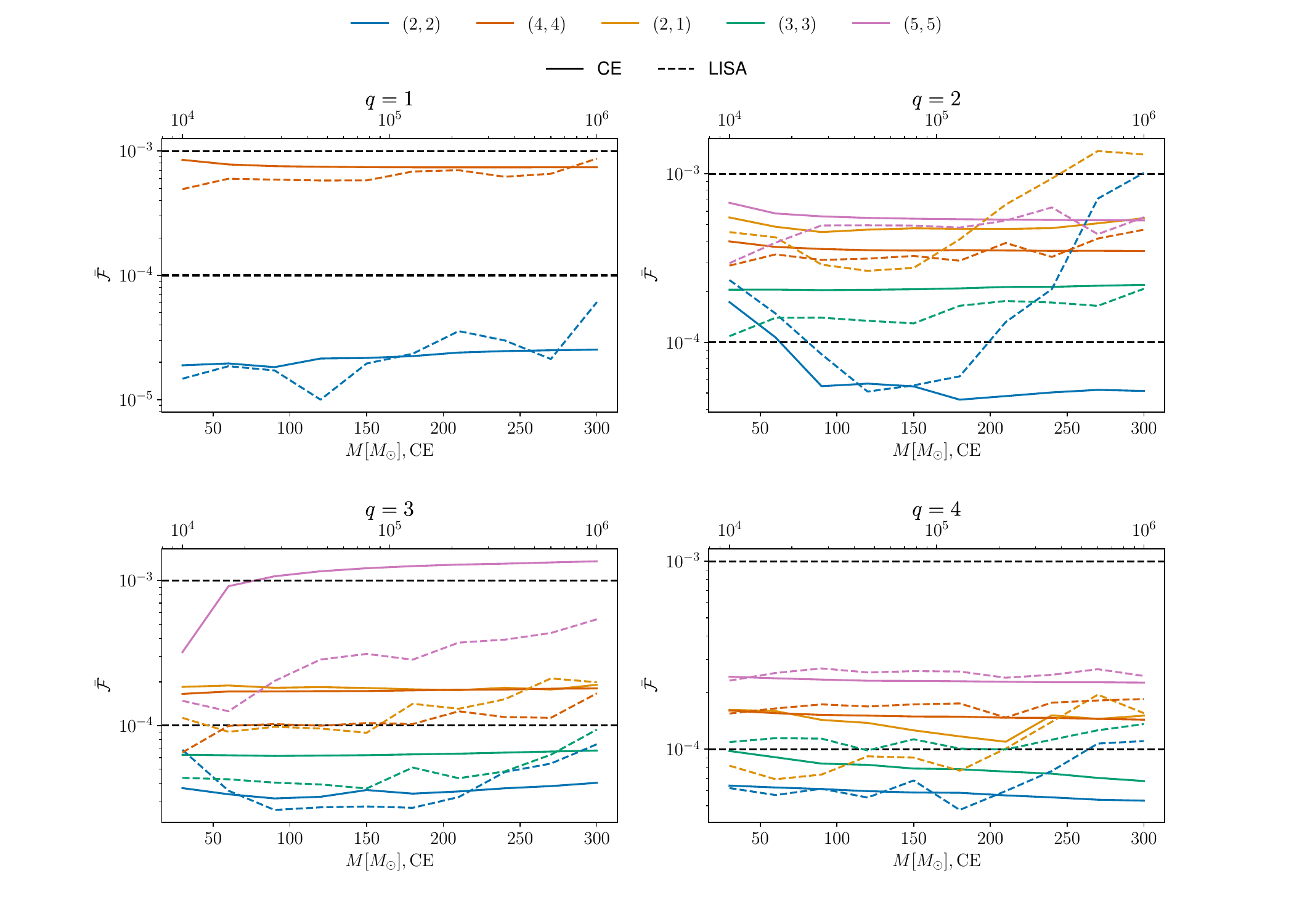}
  \caption{Single-mode mismatch between \gra{} and SXS waveforms for the $(2, 2), (2, 1), (3, 3), (4, 4), (5, 5)$ modes obtained with the CE
  (straight lines) and LISA (dashed lines) noise curves at varying mass ratios. 
  In both cases mismatches typically lie below the $10^{-3}$ threshold, with the only exception being the $(5,5)$ mode for $q=3$ and the 
  $(2,1)$ mode for $q=2$.
  Overall, the agreement between the two codes is of accuracy comparable to (or better than) current
  \ac{nr}-informed waveform models.
  }  
  \label{fig:sxs_mismatch_singlemode}
\end{figure*}

\begin{figure}[t]
  \includegraphics[width=\linewidth]{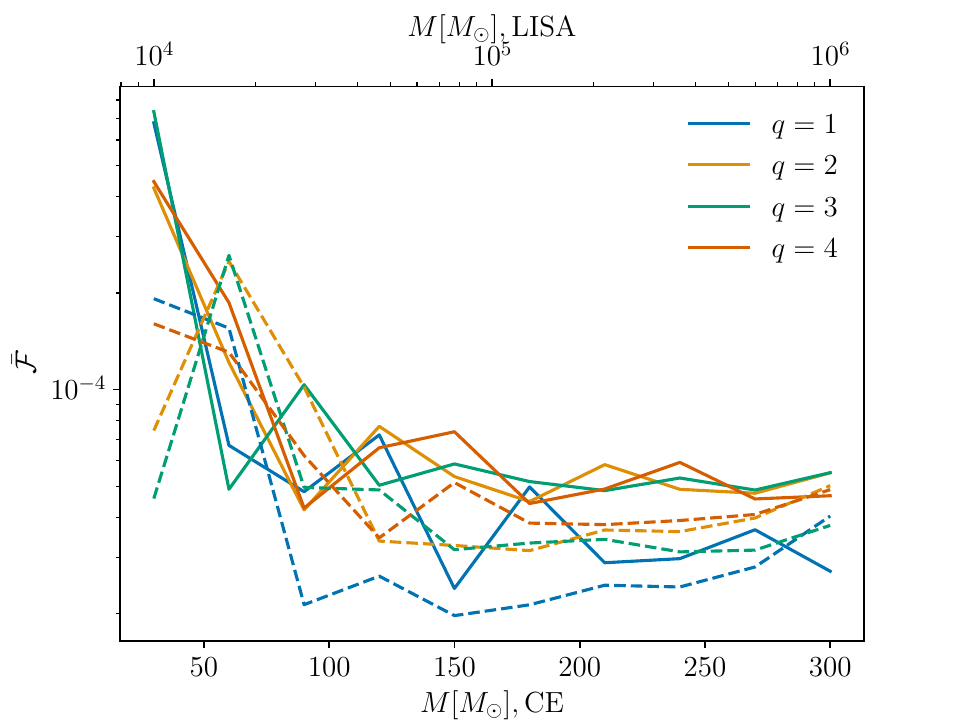}
  \caption{Mismatch between \gra{} and SXS waveforms computed using polarizations constructed from the $(2, 2), (2, 1), (3, 3), (3, 2), (4, 4), (5, 5)$ modes.
  The mismatches are averaged over four values of effective polarization angle and four values of reference phase of the target waveform.
  The agreement between the two codes is of accuracy comparable to or higher than current NR-informed waveform models for both
  the CE (straight lines) and LISA (dashed lines) noise curves, 
  and always below the $10^{-3}$ threshold.}
  \label{fig:sxs_mismatch_hms}
\end{figure}

We now compare our waveforms with equivalent simulations from the SXS catalog~\cite{Buchman:2012dw,Blackman:2015pia,Boyle:2019kee}.
In particular, we show the comparison against simulations \texttt{SXS:BBH:1155},
\texttt{SXS:BBH:1167}, \texttt{SXS:BBH:1179} and \texttt{SXS:BBH:0167}. 
While we report comparisons for these specific cases, we have also compared with a broader set of SXS waveforms, 
whose initial configurations match our waveforms.
These simulations are chosen based on their low initial eccentricity
and high resolution, making them suitable for informing waveform models~\cite{Albertini:2021tbt}.
For each SXS simulation, we employ the second extraction level and the highest resolution available.
To compare these waveforms, we first use \eq{eq:phasing} to align them, and
find $\Delta\phi_{22}$ and $\Delta t$.
Then, the other modes are aligned by employing 
the same $\Delta t$ and appropriately rescaling $\Delta \phi_{22}$:
\begin{equation}
\Delta\phi_{\ell m} = \Delta\phi_{22} \frac{m}{2} \, .
\end{equation}

\fig{fig:sxs_22_allq} shows the comparison between the $(2,2)$ mode of our waveforms and SXS for all mass ratios considered.
The visual agreement between the two waveforms is excellent, with the phase differences throughout the whole inspiral up to merger
being consistently below $0.1$ rad. Similarly, the amplitude of the waveforms is in good agreement, 
with relative differences typically below $1\%$ and small oscillations observed during the inspiral phase due to some 
residual eccentricity in the \gra{} simulations.

\fig{fig:sxs_td_q2}, instead, shows the comparison between the $q=2$
\gra{} and SXS simulations 
for the $(\ell, m) = \{(2,2), (2,1), (3,3), (3,2), (4,4), (5,5)\}$ modes.
Similar plots for the other mass ratios are available in \app{app:gra_vs_sxs}.
Again, we find good agreement between the two simulations, with the phase differences
at merger remaining at the sub-radian level for all modes investigated. The largest discrepancies are observed for the
$(3,2)$ mode, with amplitude relative differences reaching ${\sim}20\%$.
While it is difficult to ascertain the exact cause of these differences, we note that--as shown in
Fig.~9 of \rf{Ferguson:2023vta}--that the $(3,2)$ mode is rather sensitive to numerical errors due to the center of mass drift of the binary. This effect
is not corrected in the \gra{} simulations presented here. Additionally, $\psi_{32}$ is affected by large numerical noise, which may also contribute to the 
observed inconsistencies.
Overall, we find that while the dominant quadrupole modes agree to within a few tens of a radian to the SXS waveforms, 
the computation of precise higher modes is more challenging and may require further investigation, higher resolutions, or 
improved numerical techniques.

To complement the time-domain comparison, we also quantify the discrepancy between the two waveform
catalogs by computing mismatches. We compute them between single modes only, as well as for
waveform polarizations constructed using
the $(\ell, m) = \{(2,2), (2,1), (3,3), (3,2), (4,4), (5,5)\}$ modes, fixing the inclination angle to $\iota = \pi/3$.
We use \eq{eq:mismatch} to compute the mismatches as the output is independent of the sky position of the source.
When higher modes are included, we average the obtained mismatches over four values of effective polarization angle $u$.
and four values of reference phase of the target waveform. 

Results for single mode mismatches are shown in \fig{fig:sxs_mismatch_singlemode}.
When the CE noise curve is employed, mismatches for all modes are always below the $10^{-3}$ threshold,
with the notable exception of the $(5,5)$ mode for the $q=3$ case.
A clear hierarchy can be observed, with the $(2,2)$ mode showing the best agreement ($\bar{\mathcal{F}} < 10^{-4}$) for all systems considered, 
followed by the $(4,4)$, $(3,3)$, $(2,1)$ and $(5,5)$ modes.
When the LISA noise curve is used results are comparable to those previously discussed, although a slight worsening
of the agreement is observed at higher masses.
\fig{fig:sxs_mismatch_hms}, instead, shows the mismatches computed with
polarizations constructed using the $(\ell, m) = \{(2,2), (2,1), (3,3), (3,2), (4,4), (5,5)\}$ modes.
The initial frequency chosen for the mismatch calculation is $f_0 = 1.25 \times \omega_{22}5/(4\pi)$,
to ensure that we are focusing on a frequency region where all modes are present.
The mismatches found always lie comfortably below the $10^{-3}$ threshold for both the CE and LISA noise curves.

These figures indicate that the agreement between the two codes is sufficient for the purposes of current generation observations, 
and comparable to or better than the accuracy of current state-of-the-art waveform models $(\bar{\mathcal{F}} \sim 10^{-4}-10^{-3})$.
However, with next generation detectors in mind, further improvements--especially in the computation of higher modes--may be necessary. 
\section{Conclusion and discussion}
\label{sec:discussion}

Using the \gra{} code, we simulated a set of non-spinning \ac{BBH}
systems with mass ratios of 1, 2, 3, and 4, thereby
creating the first \gra{} catalog. 
We compute \ac{CCE} strains at null infinity $\mathcal{I}_+$ using the \pit{} code.
Additionally, we study their phase and amplitude convergence with respect to \pit{}'s
resolution, the world tube extraction radius, and \gra{}'s resolution.
We find that the phase evolution shows a clean 
second-order convergence
with respect to \pit's resolution as well as the world tube extraction
radius--for the dominant $(2,2)$ mode.
Moreover, the amplitude of \ac{CCE} strains for the $q=$ 2, 3, and 4 runs
shows at least a second-order convergence for both \pit's 
resolution and the world tube extraction radius,
but for $q=1$, it exhibits a first-order convergence 
with respect to \pit's resolution and no convergence 
with respect to the world tube extraction radius.
Regarding the phase convergence with respect to \gra{}'s resolution,
$q=1$ and $q=2$ \ac{BBH} runs demonstrate fourth-order and second-order
convergence, respectively; 
however, the $q=3$ and $q=4$ run 
do not show any convergence in this test. 
Similarly, the amplitude for $q=1$ and $q=2$ cases shows 
fourth-order convergence and no convergence for $q=3$ and $q=4$ runs.

Accounting for all sources of error from \pit{} and \gra{} codes, we find
that for the $q=1$ binary, the phase uncertainty is of the order of
$10^{-2}$ radians during the inspiral and peaks at ${\sim}0.1$ radians at
merger. The amplitude error similarly grows from $1\%$ during the
inspiral to up to $10\%$ at merger. In either cases, the uncertainty is
dominated by the extrapolation of the waveforms to $\mathcal{I}_+$ using
\pit{}. 
The error budget for phase and amplitudes for the other binaries
is about a factor two smaller and, for $q=2$ and $q=3$, is also dominated
by \ac{CCE}. The $q=4$ runs, are the only one for which the
finite-resolution error is comparable --- although smaller than the
\ac{CCE} error.

Our self-mismatch study for the $(2,2)$ mode of the \ac{CCE} strains for
mass ratios of 1, 2, 3, and 4,
using the LISA noise curve and total masses ranging from
$(10^4, 10^6) M_\odot$, yields a mismatch of approximately $10^{-12}$ to
$10^{-11}$ for a \ac{CCE} inner radius of 50.
Calculating the self-mismatch with the same settings but with inner radius of
100, we obtain a mismatch of approximately $10^{-5}$ to $10^{-4}$.
Unfortunately, the differences between waveforms at $\mathcal{I}_+$ obtained by
running \pit{} for different inner radii are substantial. 
This is an
unexpected finding, which indicates that even though the intrinsic error
in the \gra{} simulations is very small, the overall quality of our
waveform is somewhat compromised.
These are nevertheless among the most
accurate puncture-based waveforms for \ac{BBH} to date.
Additionally, our preliminary results using \texttt{SpECTRE CCE}~\cite{Moxon:2021gbv}
indicate that the \ac{CCE} strains obtained using different world tube radii, namely, 
50 and 100, show greater consistency in terms of error and mismatch. 
Consequently, we intend to conduct further experiments with
\texttt{SpECTRE CCE} in future work.

We compared our results with those from the SXS catalog and confirmed
that our waveforms are in agreement with those produced by the SXS
collaboration, with typical phase differences below $0.1$ radians throughout
the inspiral and merger phases and \textit{maximum} mismatches of the
order of $10^{-3}$ for both the \ac{ce} and LISA noise curves.  These
numbers indicate that our waveforms are in good agreement with those
produced by the SXS collaboration for the considered number of cycles. 
This validation process is essential for establishing the credibility 
of our results and ensuring that they align with existing data from other 
established sources. Our work also provides another independent confirmation 
of the accuracy of the SXS waveform catalog, which is widely used for \ac{GW} 
waveform modeling and data analysis applications. 
That said, the differences between our waveforms and those from the SXS 
collaboration are significantly larger than the uncertainties we have estimated for our data. 
These inconsistencies could be the result of gauge or initial data differences,
or due to some other, unidentified source of systematic uncertainty in
our waveforms or in the SXS catalog. Additional work, carefully cross
validating \gra{} with SXS's \texttt{SpEC} \cite{Boyle:2019kee} and
\texttt{SpECTRE} \cite{Kidder:2016hev,Lovelace:2024wra} codes, 
is needed to understand the origin of these discrepancies.



Our work demonstrate that finite-differencing moving-puncture methods can
produce highly accurate \ac{BBH} simulations, with quality comparable to
those obtained with spectral methods. On the other hand, we have
identified a new source of systematic uncertainties in \ac{nr} data in
the commonly adopted method to extrapolate waveforms to future
null-infinity via \ac{CCE}. This dominates the error budget for our
waveforms. Future work will be dedicated to address this source of error,
in particular, we plan to interface \gra{} with the publicly available
\ac{CCE} in \texttt{SpECTRE} \cite{Moxon:2021gbv}. We also plan to extend
the \gra{} catalog with the inclusion of higher mass ratios, eccentric,
and spinning \ac{BBH} configurations.  Finally,  additional work is
needed to understand the discrepancies between our results and those of
the SXS catalog, starting with a more detailed investigation of the
effect of eccentricity in the \gra{} simulations and the impact of the
center of mass drift.

\begin{acknowledgments}
The authors would like to thank A.~Nagar and S.~Albanesi for valuable discussions
thoughout the development of this project.
This work was supported by NASA under Award No. 80NSSC21K1720.
RG acknowledges support from NSF Grant PHY-2020275 (Network for Neutrinos, Nuclear Astrophysics, 
and Symmetries (N3AS)).
KC acknowledge the support through NSF grant numbers PHY-2207638, AST-2307147, PHY-2308886, and PHY-2309064.
SB and BD knowledge support by the EU Horizon under ERC Consolidator Grant, no. InspiReM-101043372.
  
Simulations were performed on Frontera (NSF LRAC allocation PHY23001), 
and on the national HPE Apollo Hawk at the High Performance 
Computing Center Stuttgart (HLRS).
The authors acknowledge HLRS for funding this project by providing
access to the supercomputer HPE Apollo Hawk under the grant number
INTRHYGUE/44215 and MAGNETIST/44288.
\end{acknowledgments}

\appendix

\section{Finite radius extraction method}
\label{sec:fre_method}

In the \ac{FRE} method, the $\mpsi$ modes are extracted at a specific radius $R$ to construct the Newman-Penrose scalar $\Psi_4$ as follows~\cite{Newman:1961qr, Bruegmann:2006ulg}:
\be
  \Psi_4 = 
  \sum_{\ell=2}^{\infty}\sum_{m=-\ell}^\ell \mpsi(t)\;
  {}_{-2}Y^{\ell m}(\vartheta,\varphi),
  \label{eq:psi_lm}
\ee
in which ${}_{-2}Y^{\ell m}(\vartheta,\varphi)$ is the spin-2 weighted spherical harmonic. $\Psi_4$ is also related to
$h_+$ and $h_{\times}$ via \eq{eq:psi4_news}.

To obtain \ac{GW} strains at null infinity,
we first use the following formula to construct $\mpsi$ at null infinity
\cite{Lousto:2010qx,Kiuchi:2017pte}
\begin{align}
\label{eq:psi_extrapolation}
\lim_{r \rightarrow \infty} r \psi_4^\lm \simeq & A 
\bigg( \bar{r} \psi_4^\lm- \frac{(\ell-1)(\ell+2)}{2 \bar{r}} \int dt \, \bar{r} \psi_4^\lm \bigg)
\\ \nonumber
    + & \mathcal O(R^{-2})
\end{align}
where $A=1-2M/\bar{r}$ and $\bar{r}=R (1+M/(2R))^2$.
We then use either a time domain 
integration or an \ac{FFI} method~(see \app{sec:ffi_method})
to obtain \ac{GW} strains at null infinity.

\section{Fixed frequency integration}
\label{sec:ffi_method}

The time domain integration method is error prone, involving the subtraction of
various polynomials during each time integration step, which can potentially
impact the waveform content.
To avoid this, we use the \ac{FFI} method~\cite{Reisswig:2010di}.

\begin{figure}[hbt!]
\centering
\includegraphics[width=1\linewidth,clip=true]{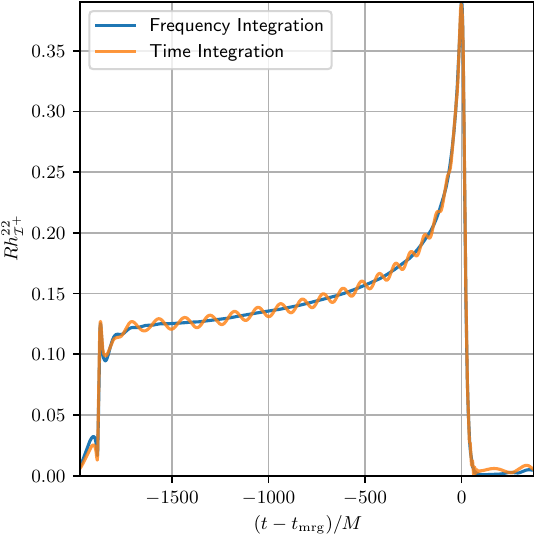}
\caption{%
Strains at null infinity~($\mathcal{I^{+}}$) 
for the $q=1$ system. 
We use two different integration methods to compute 
$r\mpsitt$ at $\mathcal{I^{+}}$, namely \ac{FFI} (blue) and
time domain integration (orange).
Given $r\mpsitt$ at $\mathcal{I^{+}}$,
we use \ac{FFI} to calculate the strains.
We observe that time domain integration gives rise to nonphysical modulations in
the middle of the strain and values larger than the frequency integration
in the tail.
}

\label{fig:int_vs_ffi_extrap}
\end{figure}

During the \ac{FFI} process, preserving the waveform's physical frequency range is essential. To achieve this, the following cutoff frequency:
\be
f^{\mrm{cutoff}}(m) = \frac{\Omega_{\mrm{orb}}}{3\pi} \max(1,m),
\label{eq:cutoff_freq}
\ee
is used for each $m$ mode. 
Here, $\Omega_{\mrm{orb}}$ is the orbital angular velocity calculated
using the Newtonian approximation $\Omega_{\mrm{orb}}=\frac{(q+1)^2}{q D}P_y$,

Finally,
it is worth mentioning that to compute the second term of 
\eq{eq:psi_extrapolation} we use \ac{FFI}.
Time-domain integration introduces nonphysical modulations
in strain amplitude, and post-merger waveforms
do not converge to zero as seen in \fig{fig:int_vs_ffi_extrap}. 
%

\section{FRE convergence}
\label{sec:fre_convergence}
%
\begin{figure*}[t]
\centering
\includegraphics[width=1\linewidth,clip=true]{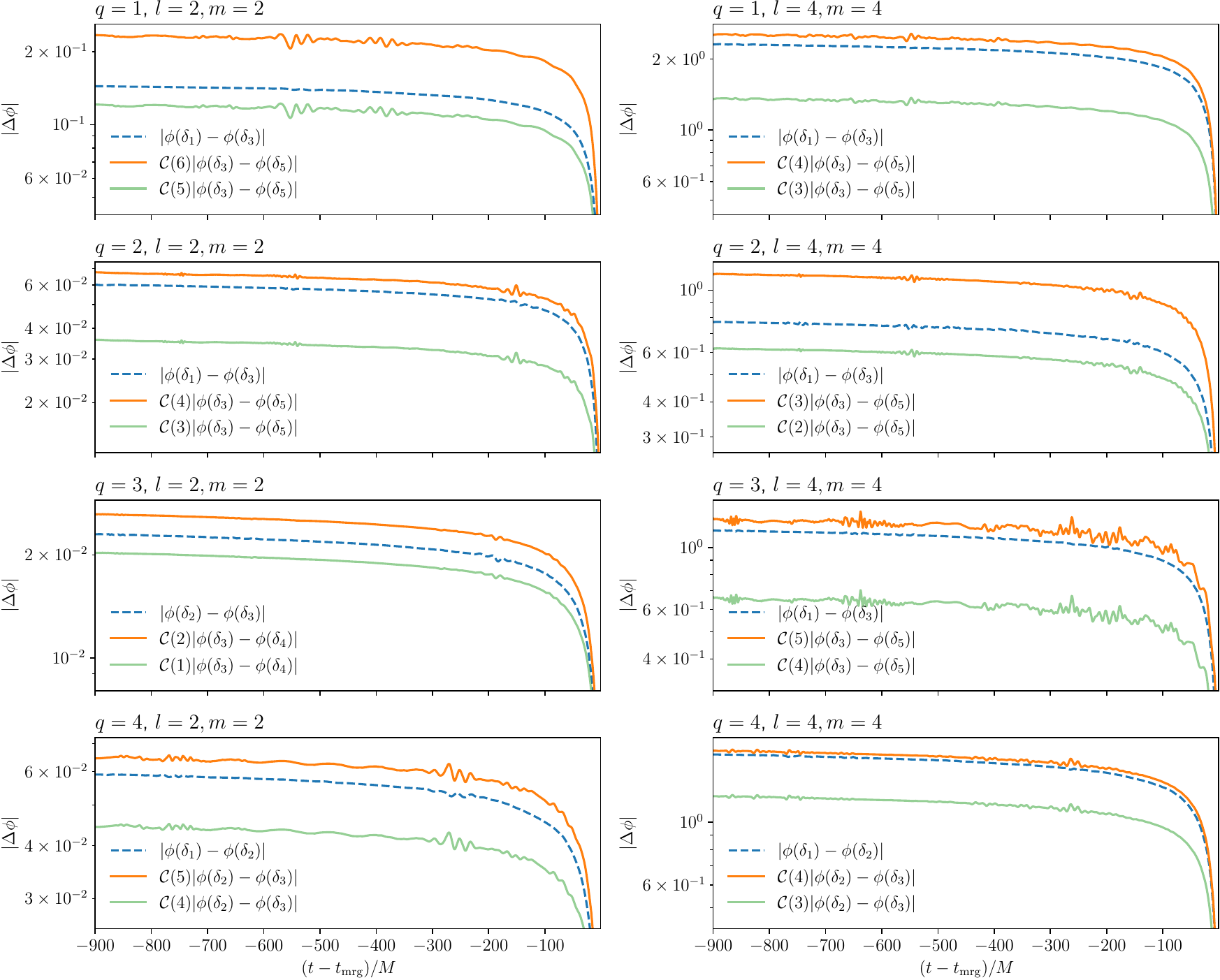}
\caption{%
Convergence study of strains using the \ac{FRE} method.
Here, $\delta_i^{-1} \in \left \{128,192,256,320,384 \right \}$,
for $i=1,2,3,4,5$.
We extrapolate $\mpsi$ to null infinity using \eq{eq:psi_extrapolation}.
The pertinent strain is computed via \ac{FFI}.
Left column shows $l=2,m=2$ mode and right column shows $l=4,m=4$ mode for
different mass ratios. 
The location of dashed line with respect to colored solid lines indicates the order
of convergence. 
}
\label{fig:conv_test_fre_res_h_from_psi4}
\end{figure*}
\begin{figure*}[t]
\centering
\includegraphics[width=1\linewidth,clip=true]{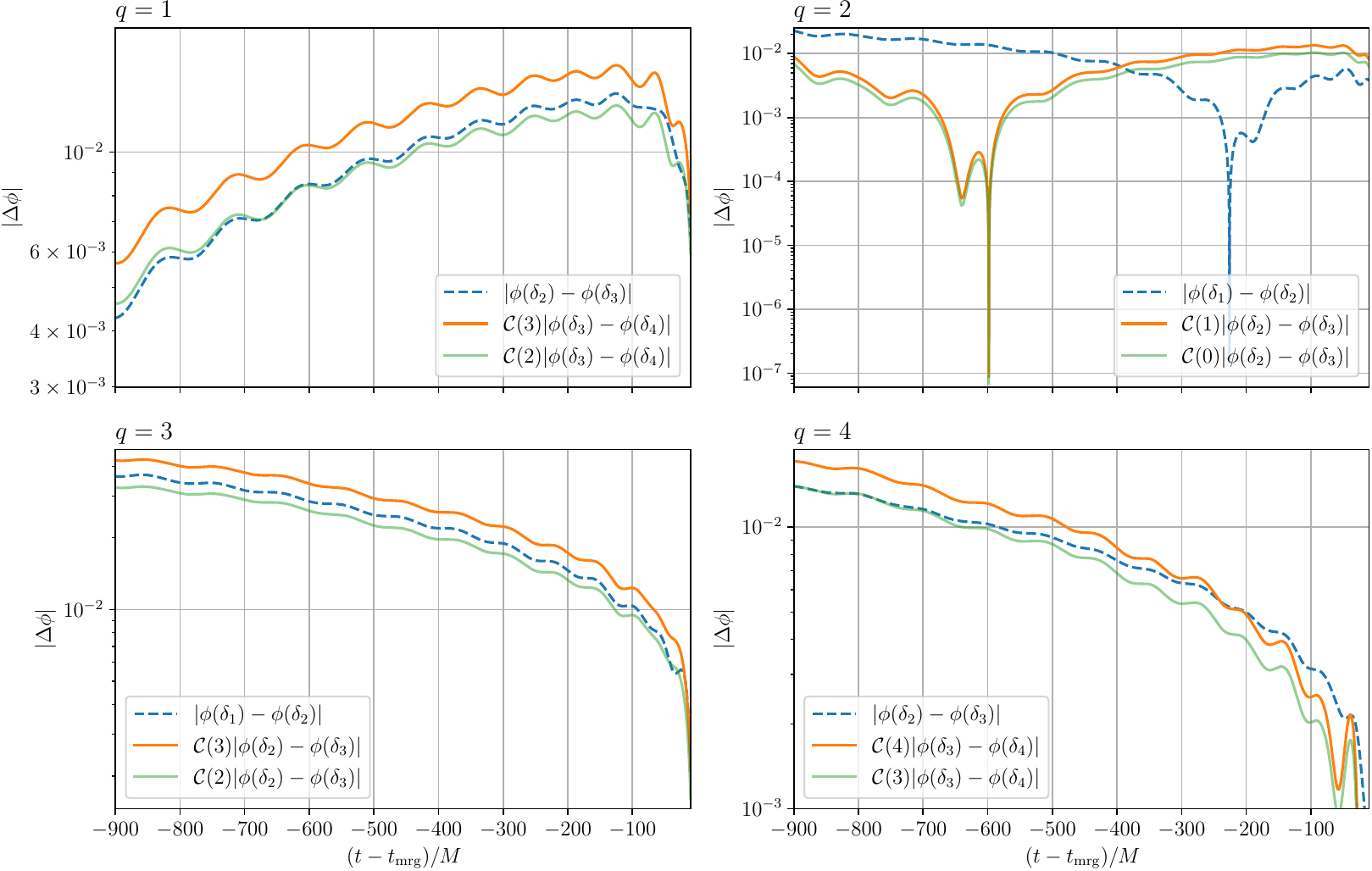}
\caption{%
Radius convergence of the \ac{FRE} (2,2) mode for different mass ratios.
Here, $\delta_i^{-1} \in [60,80,100,120]$.
We observe convergence for all mass ratios, except for the $q=2$ case.
}
\label{fig:conv_test_fre_radius_psi4}
\end{figure*}
\begin{figure*}[hbt!]
\centering
\includegraphics[width=1\linewidth,clip=true]{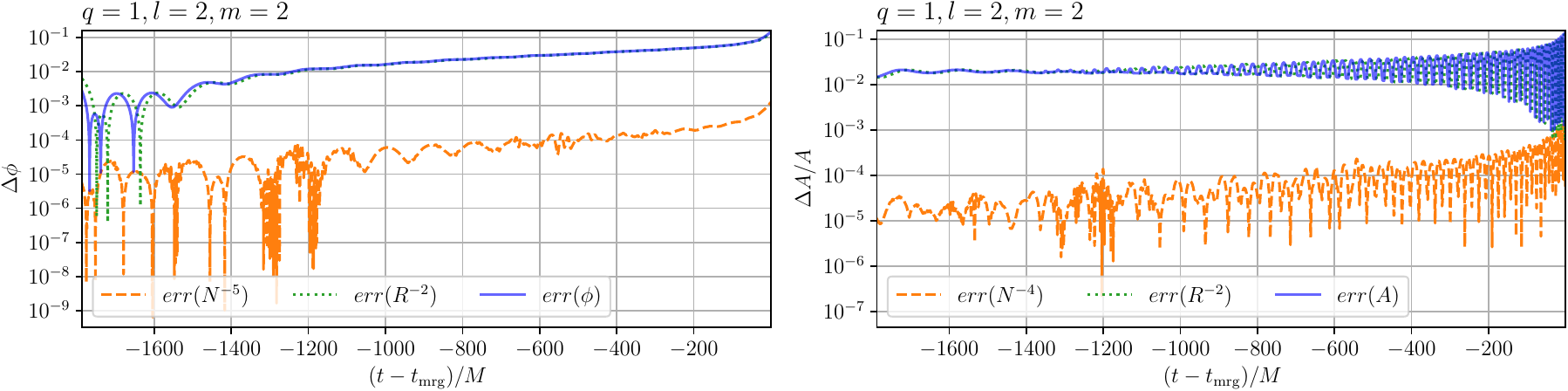}
\includegraphics[width=1\linewidth,clip=true]{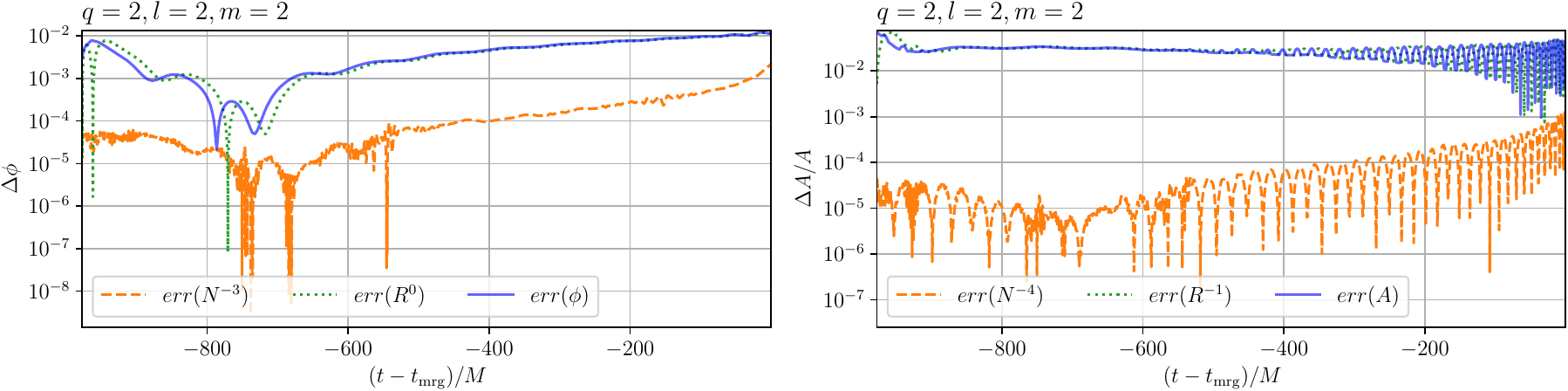}
\includegraphics[width=1\linewidth,clip=true]{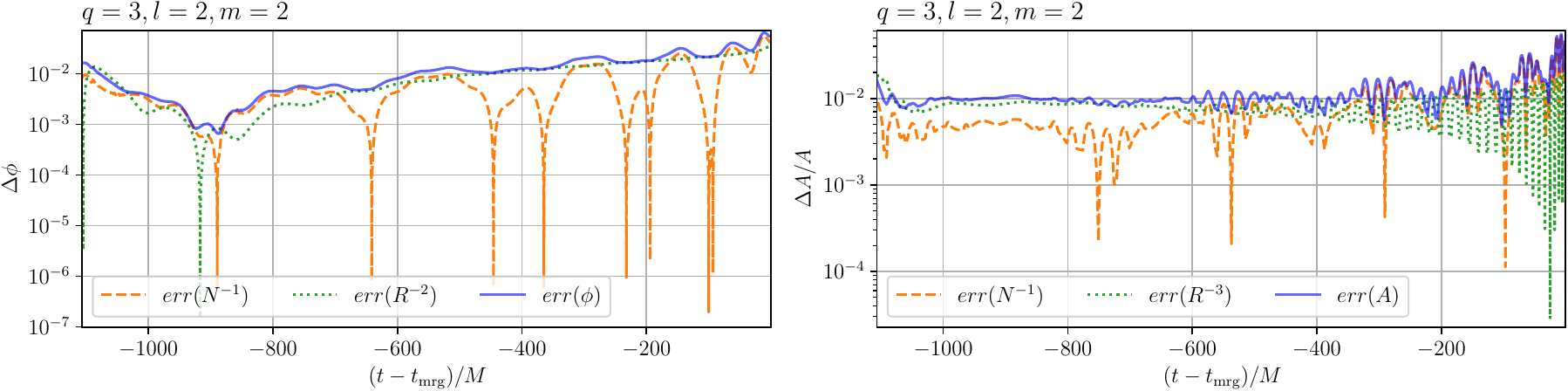}
\includegraphics[width=1\linewidth,clip=true]{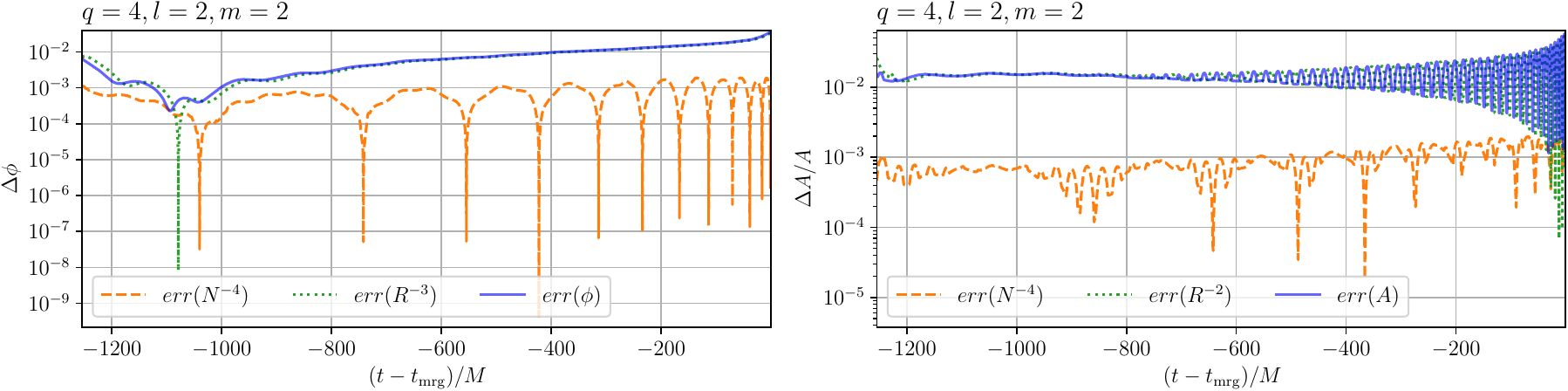}
\caption{%
Sum of errors in the \ac{FRE} strains for the (2,2) mode.
$err(N^{-p})$ denotes the error of the highest resolution run with
the extraction radius of 100 in the table~(\ref{tab:bbh_runs}).
$p$ indicates the convergence order.
Similarly, $err(R^{-p})$ represents the error for the radius 100, 
where to find the error we use radii 80 and 100 in \eq{eq:trunc_error}.
The total error $err$ for $\phi$ as well as the relative amplitude $\Delta A/A$
is computed by
$err = (err^2(N^{-p}) + err^2(R^{-p}))^{1/2}$.
}
\label{fig:err_budget_fre_h_from_psi4}
\end{figure*}

This section quantifies the numerical errors in the \ac{GW} strain obtained using the \ac{FRE} method due to resolution and radius extraction at null infinity. We use the same procedure outlined in Sec.~\ref{sec:cce_convergence}, and the results of the convergence analysis for the phase is shown in \fig{fig:conv_test_fre_res_h_from_psi4}

%

We extract the waveforms at $R=100$ and employ three distinct
resolutions from the table~(\ref{tab:bbh_runs}). Each row in
\fig{fig:conv_test_fre_res_h_from_psi4} corresponds to a specific mass ratio, while
each column corresponds to a specific mode.
Focusing on the dominante $(2,2)$ and $(4,4)$ modes, we use the method described 
in Sec.~(\ref{sec:conv_test}), and we find convergence behavior in all cases.
It is important to note that we anticipate a convergence order in resolution 
between 4 and 6 when the simulation is within the convergence regime~(see Sec.~\ref{sec:scheme}). 
As resolution increases, results should approach the lowest convergence order of 4, 
corresponding to the 4th order Runge-Kutta time integrator.
Until this convergence rate is confirmed, the estimated order can vary
between 4th and 6th order, as observed convergence can vary with resolution.
However, identifying the same consistent convergence behavior across all simulations 
is complicated. A puncture solution of Einstein's equations for \ac{BBH} systems is not 
a regular solution and typically exhibits only $C^2$ continuity 
at the punctures~\cite{Ansorg:2004ds}. 
Additionally, the use of different resolutions for runs with varying mass ratios 
further complicates this analysis.

Similarly, we choose the highest resolution and perform the analysis 
for different extraction radii to determine the convergence order for the
phase and amplitude of the \ac{FRE} strains.
The outcomes of this analysis are illustrated
in \fig{fig:conv_test_fre_radius_psi4}. Across all cases, except
$q=2$, we observe that strains converge with respect to the extraction radius\footnote{While
for $q=2$ run we see no convergence in phase with respect 
to the extraction radii, we observe a 2nd order convergence in amplitude.}.

After determining the convergence order, we quantify the leading order 
error term for the highest available resolution waveform with $R=100$.
To assess the resolution error, we align this waveform with 
the second-highest resolution one using \eq{eq:phasing} and 
apply equation \eq{eq:trunc_error}. 
Similarly, we use the highest resolution waveform 
extracted at radius 100 and 80 for the extraction radius error, and then employ equation \eq{eq:trunc_error}.

\fig{fig:err_budget_fre_h_from_psi4} summarizes the error analysis results.
The term $err(N^{-p})$ represents the error associated to resolution of
\gra{}, while $err(R^{-p})$ denotes the error arising form the extraction
radius. Here, $p$ indicates the order of convergence found in each
convergence study.
We observe 
that the dominant error arises from the extraction radius,
and the total error increases as we approach the merger. 
The largest errors are observed for the $q=1$ case, where the total error
reaches ${\sim} 0.1$ radians, and the relative amplitude difference becomes
${\sim} 10\%$. Higher mass ratio runs show smaller errors, with $q=2$
approaching $\Delta A/A \sim 5\%$ and $\Delta\phi \sim 0.01$ radians.
For the $q=3$ and 4 systems, the maximum total error is 
${\sim} 0.05$ radians in phase, and ${\sim} 5\%$ in the relative amplitude
difference. 
%



%

\section{FRE vs CCE}
\label{sec:fre_vs_cce}
%
\begin{figure*}[t]
\centering
\includegraphics[width=1\linewidth,clip=true]{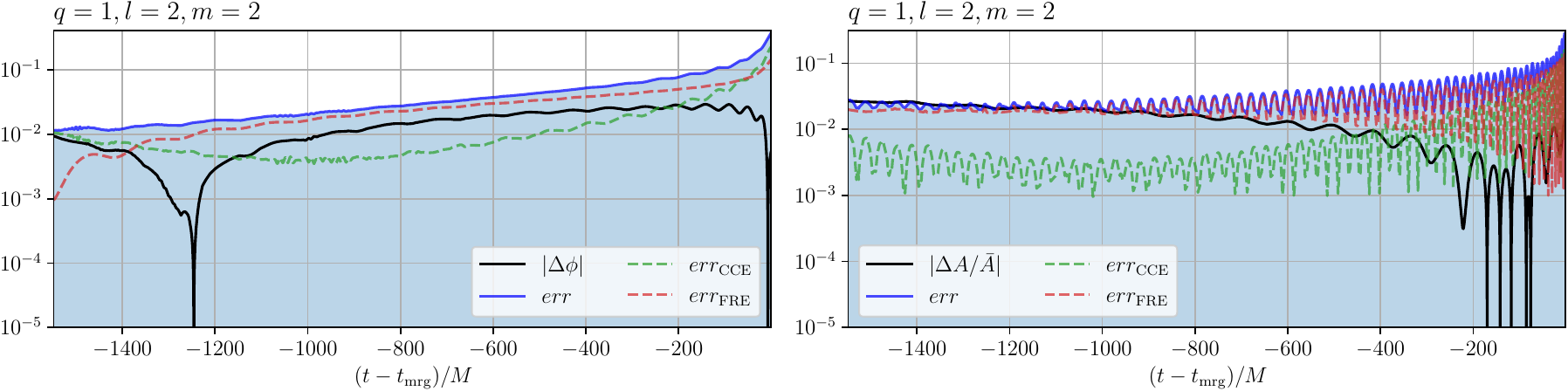}
\includegraphics[width=1\linewidth,clip=true]{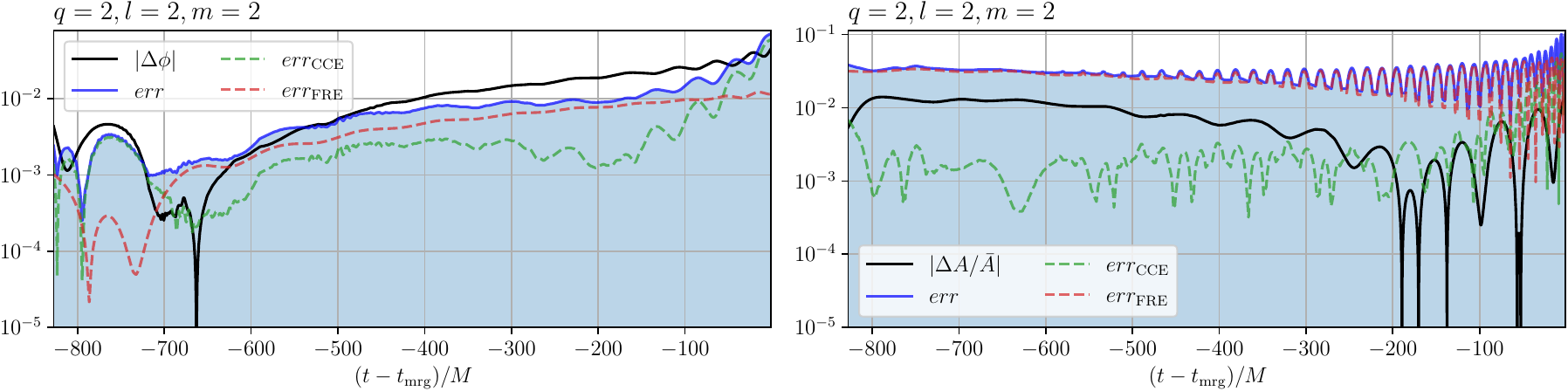}
\includegraphics[width=1\linewidth,clip=true]{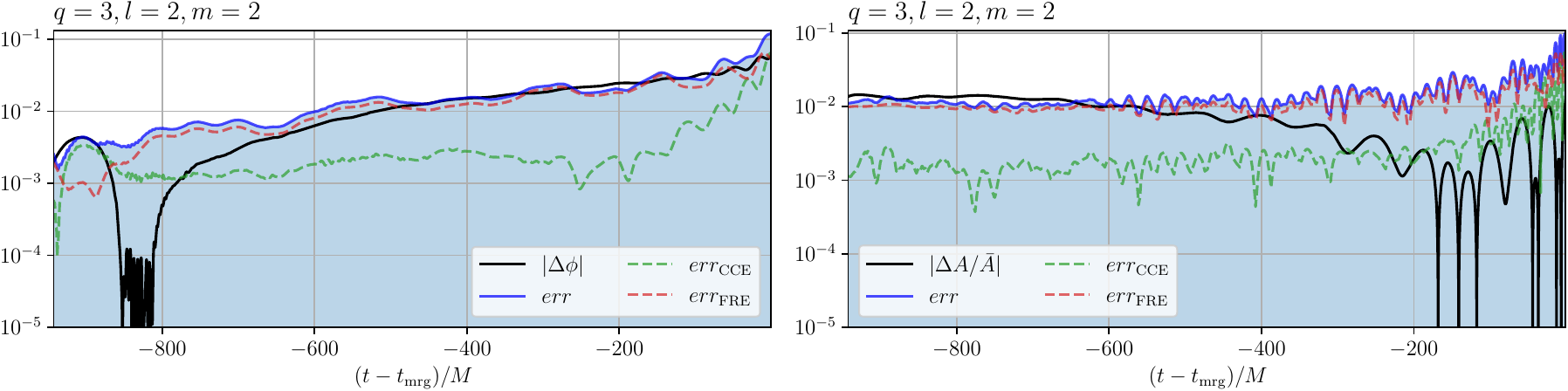}
\includegraphics[width=1\linewidth,clip=true]{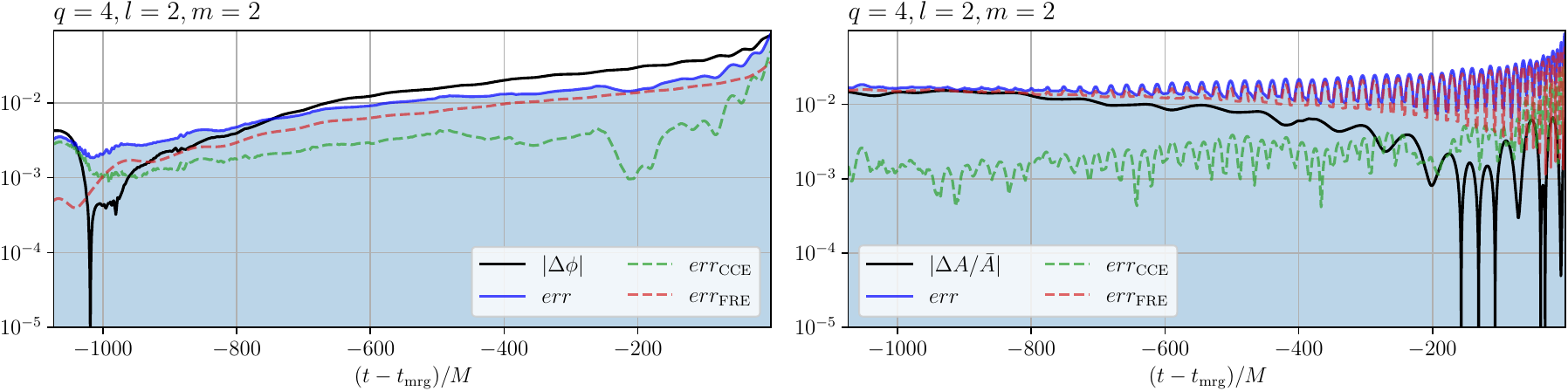}
\caption{%
Difference between \ac{FRE} and \ac{CCE} strains.
Left columns show the difference between the phase of \ac{FRE} and \ac{CCE}
strains, $|\Delta \phi| = |\phi_{\mrm{CCE}} - \phi_{\mrm{FRE}}|$ -- marked by
black solid lines.
Right columns show the relative difference between the amplitudes~(the solid
black line), i.e.,
$
|\Delta A / \bar{A}| := 
\left |(A_{\mrm{CCE}} - A_{\mrm{FRE}})/\bar{A}\right|,
$ where $\bar{A}$ is the arithmetic mean of the amplitude, 
$\bar{A} = (A_{\mrm{CCE}} + A_{\mrm{FRE}})/2$.
Each row illustrates these comparisons for different mass ratio with the
highest available resolution from table~(\ref{tab:bbh_runs}).
The numerical errors for the \pit{} code~($err_{\mrm{CCE}}$), 
and the \gra{} code~($err_{\mrm{FRE}}$) are illustrated by dashed lines.
The total numerical error is marked by a blue solid line in each panel as
well as shaded under its curve for a better visualization.
We compute the total numerical error in dephasing by using 
$err = |err_{\mrm{PIT}}| + |err_{\mrm{GRA}}|$, and
for the amplitude relative difference by 
$
err = \left(|err_{\mrm{CCE}} A_{\mrm{CCE}}| + 
            |err_{\mrm{FRE}} A_{\mrm{FRE}}| \right)/ \bar{A}.
$
}
\label{fig:h_cce_vs_fre}
\end{figure*}

We investigate the discrepancies between the (2,2) modes of the \ac{CCE} and
\ac{FRE} strains for our simulations.
The left column of \fig{fig:h_cce_vs_fre} compares the phase
difference $|\Delta \phi|$ between the highest available resolution of
\gra{}~(\ac{FRE} strain with radius 100), and \pit{}~(\ac{CCE} strain with
radius 50). Similarly, in the right column, 
we compare the amplitude between the \ac{FRE} and \ac{CCE} strains using
$
|\Delta A / \bar{A}| := 
\left |(A_{\mrm{CCE}} - A_{\mrm{FRE}})/\bar{A}\right|,
$
where $\bar{A} = (A_{\mrm{CCE}} + A_{\mrm{FRE}})/2$.
%
%
The difference between the two methods, i.e., 
$\Delta \phi$ and $|\Delta A / \bar{A}|$, is shown by a black solid line.
We observe that $\Delta \phi$ increases while 
$|\Delta A / \bar{A}|$ decreases as we approach merger.
The maximum $\Delta \phi$ is ${\sim} 0.1$ radians and is found for the $q=4$
run, and occurs at the merger;
however, the other mass ratio runs have slightly smaller errors compared to $q=4$.
Lastly, $|\Delta A / \bar{A}|$, among all runs, barely reaches
$3\%$ relative difference, cf.~\cite{Albanesi:2024xus}.

Using numerical error estimates for the phase, 
obtained in Sec.~\ref{sec:cce_convergence} and Sec.~\ref{sec:fre_convergence},
we plot the total numerical error
$
err = |err_{\mrm{CCE}}| + |err_{\mrm{FRE}}|
$
in \fig{fig:h_cce_vs_fre}.
We see that in $q=$ 1 and 3 case, the total numerical error is bigger than
or comparable with $\Delta \phi$,
and hence, we can expect that the difference 
in phases of the \ac{FRE} and \ac{CCE} strains are due to the numerical errors 
contained within each code, i.e., \gra{} and \pit.
However, for $q=$ 2 and 4, we observe dephasing is greater than the numerical error, 
therefore, it suggests that we are underestimating the error for one or both
waveforms.

Similarly, we use the numerical errors of \gra{} and \pit{} codes
for amplitude found in Sec.~\ref{sec:cce_convergence} and 
Sec.~\ref{sec:fre_convergence},
to compute the total numerical error by
$
err = \left(|err_{\mrm{CCE}} A_{\mrm{CCE}}| + 
      |err_{\mrm{FRE}} A_{\mrm{FRE}}|\right)/ \bar{A}.
$
For the amplitude comparison, we see that total numerical error is greater than
or comparable to $|\Delta A / \bar{A}|$. As such,
it indicates that the observed deviation between the \ac{FRE} and \ac{CCE}
amplitudes falls within the numerical error region,
suggesting that the differences between the two methods
can be attributed to the expected numerical errors in the codes.

Lastly, when we compare the numerical errors of 
$err_{\mrm{CCE}}$ and $err_{\mrm{FRE}}$, we see that after the junk radiation
$err_{\mrm{CCE}}$ is always the smaller one, in all cases.
\section{Fourier filtering in CCE}
\label{sec:cce_fourier_filter}

\begin{figure}[hbt!]
\centering
\includegraphics[width=1.0\linewidth,clip=true]{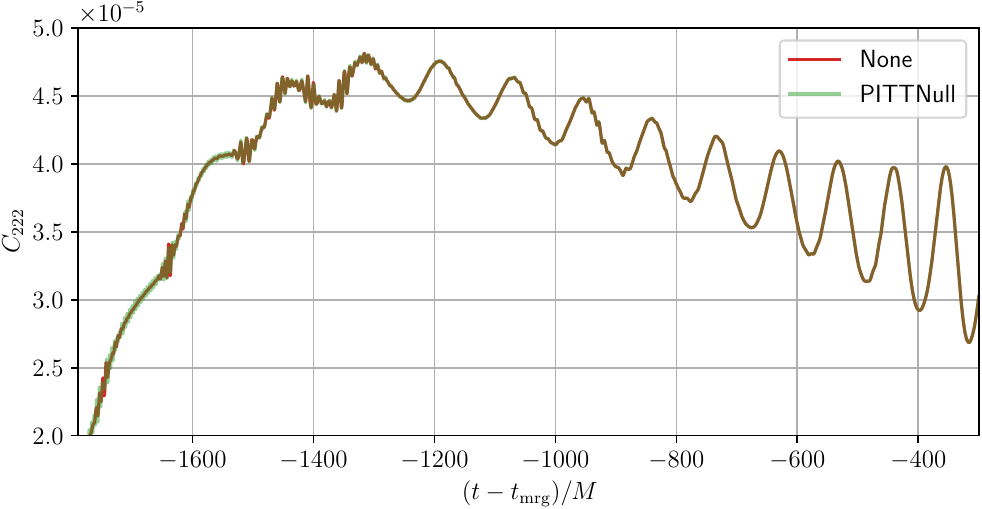}
\caption{%
The coefficient amplitude of $C_{222}$ is plotted for the spatial metric
field $g_{xx}$.
The dashed line is the original \qt{None} filtered data,
while the solid line, \pit, is the filtered data.
The coefficient does not show smooth change from one time step to the next 
at the beginning of the evolution.
In an effort to damp the noises, \cite{Babiuc:2010ze} 
drops high frequencies of $C_{klm}$ using \eq{eq:erf_filter}.
}
\label{fig:cce_coeffs_exp}
\end{figure}

At the beginning of the evolution, and sometimes even at later stages,
$C_{klm}$ in \eq{eq:cce_metric_exp} exhibits noise. 
This noise, often stemming from
numerical errors, manifests as high-frequency signals with small amplitudes,
as depicted in \fig{fig:cce_coeffs_exp}. In an effort to mitigate these
errors, \rf{Babiuc:2010ze} drops the high frequency content of the signal.
In particular, \rf{Babiuc:2010ze} employs a Fourier transformation of 
the signal ${h}(t)$ to obtain the signal in Fourier space, $\tilde{h}(k)$, 
where $k = \frac{\omega T}{2 \pi}$, 
$T$ is the total time, $\omega = \frac{d\phi(t)}{dt}$, 
and $\phi(t)$ is $\mpsi$ phase.
Then, they truncate the high-frequency noises by multiplying $\tilde{h}(k)$ with
\be
\label{eq:erf_filter}
\frac{1}{2} \left( 1-\erf \left( -(k-k_{\mrm{max}}) \right )\right),\,
k_{\mrm{max}} = \frac{\omega_{\mrm{max}} T}{2 \pi},
\ee
in which $\omega_{\mrm{max}}$ is the maximum of
$\frac{d\phi(t)}{dt}$. 
Subsequently, through an inverse transformation, they obtain a refined
strain in time.

Since Fourier transformation assumes periodicity of the wave,
a challenge is encountered when the wave is not periodic. 
To enforce periodicity, one might be inclined to apply 
windowing techniques at both ends of the data. But,
it is crucial to recognize that the data pertains
to metrics~(through $C_{klm}$) rather than $\mpsi$. 
As such, applying windowing on the data is prone 
to introducing nonphysical values for the metric. 
To address this periodicity requirement,
Ref. \cite{Babiuc:2010ze} subtracts a second-order polynomial from 
both ends of the signal, thereby imposing periodicity. 
Following the removal of higher frequencies, the
previously subtracted polynomial is reintroduced to yield the filtered data.

The comparison presented in \fig{fig:cce_coeffs_exp}
illustrates the efficacy of the filtering methodology proposed by
\cite{Babiuc:2010ze}. When compared with the unfiltered data, this
Fourier-based filtering technique demonstrates the capacity to attenuate
a portion of the high-frequency noise. 

Nevertheless, the presence of imperfect periodicity at the edges of the 
dataset introduces spectral leakage, impeding a rapid decay of Fourier coefficients. 
Consequently, when using the \pit{} code to analyze $\mpsi$ and $\mnews$, 
oscillatory patterns become apparent, particularly at the boundaries of the
data.

\begin{figure}[hbt!]
\centering
\includegraphics[width=1\linewidth,clip=true]{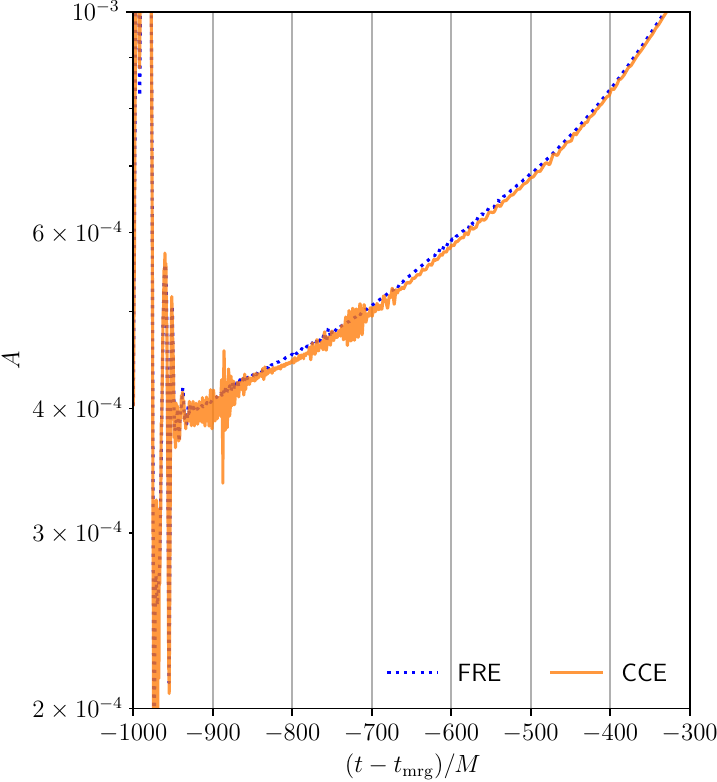}
\caption{%
The \ac{FRE} method versus the \ac{CCE} method. The \ac{FRE} method
seems smooth while \ac{CCE} exhibits oscillations, specially at the beginning
of the run.
}
\label{fig:fre_vs_raw_cce_psi4_amp}
\end{figure}

To assess the impact of noise present in the coefficients,
we compare the \ac{FRE} and \ac{CCE} $\mpsi$. The comparison is
illustrated in \fig{fig:fre_vs_raw_cce_psi4_amp}. 
The results indicate that the \ac{FRE} method yields smoother data.
Nevertheless, as shown in \app{sec:fre_vs_cce}, the \ac{FRE} strains exhibit larger
uncertainties compared to the \ac{CCE} strains.

In an effort to enhance the filtering method further, a strategy is devised to
enforce perfect periodicity on the data without introducing any nonphysical
data points. This is achieved by smoothly extending the data from both
ends using an odd reflection technique. Essentially, an odd transformation
is applied with respect to each endpoint of the waveform to extend it
symmetrically. Subsequently, a window function is used to nullify
the waveform values in the artificially extended regions. Following this
step, the extended data undergoes truncation to retain only the physically
meaningful portion.

\begin{figure}[hbt!]
\centering
\includegraphics[width=1.0\linewidth,clip=true]{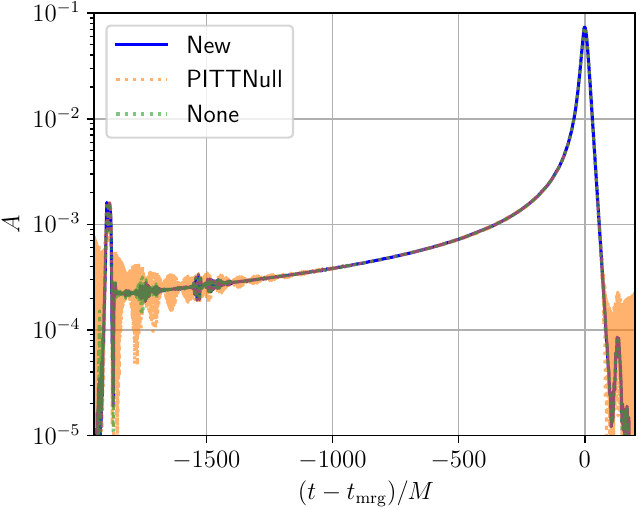}
\caption{%
Comparison of $\psi_4^{22}$ amplitudes, for $q=1$ \ac{BBH} system,
between original \pit{} filter developed in \cite{Babiuc:2010ze}
and the \qt{New} filter developed in this work, 
as well as no filtering~(\qt{None}).
The \pit{} filter shows strong oscillations at the tails due to non-perfect
periodicity.
The New filter displays slight oscillation at the tails due to
imposing extended periodicity; the amplitude of these oscillations are smaller than
the \pit{} filter. 
Comparing the \pit{} filter with the \qt{None} filter, the
\pit{} filter amplifies the noises at early times while the \qt{New} filter
damps the noises.
}
\label{fig:filters_nonfilter_psi4_amp_comparisons}
\end{figure}

It is important to acknowledge that as a result of extending the data
and applying filtering, the tails of the data exhibit some wave
patterns. 
Despite the enhancements made to the Fourier filtering technique, resulting
in smoother $\mpsi$ and $\mnews$ data, residual noise still persists. The
comparison presented in \fig{fig:filters_nonfilter_psi4_amp_comparisons}
highlights the amplitude disparities between the original filtered data,
the improved filtered data, and the unfiltered dataset.

The original filtering process of \rf{Babiuc:2010ze} 
which applied on signals that are not perfectly periodic,
while marginally smoothing the $C_{klm}$
data, introduces pronounced oscillations after using the filtered $C_{klm}$
in the \pit{} code, particularly at the waveform tails, 
and hence it introduces additional noise compared to the original unfiltered data.
As such, the original filter is less favorable for filtering purposes.
Conversely, the improved ``New'' filter, which imposes perfect periodicity, 
demonstrates a reduction in noise
levels compared to the original data. However, due to the imposition of
periodicity, the improved filtering method introduces minor oscillations
only at the waveform tails.

Although the improved filtering method shows effectiveness in reducing
noise at the initial stages of the simulations, however it may introduce modulations at
later times.

Given the challenges encountered with filtering methods, including data alterations and waveform noises' persistence, the decision
is made to forego filtering altogether. Additionally, using the \ac{FFI} method
on $\mpsi$ to obtain strains implies filtering large frequency noises that
appear in $\mpsi$. As we observe in \fig{fig:h_catalog}, the resultant
strains are mostly free of noises.

With filtering not used, the focus
shifts to determining the most suitable extraction radius for obtaining the
boundary condition for the \pit{} code. This choice becomes pivotal in ensuring the
accuracy and reliability of the data extraction process.

\begin{figure}[t]
\centering
\includegraphics[width=1.0\linewidth,clip=true]{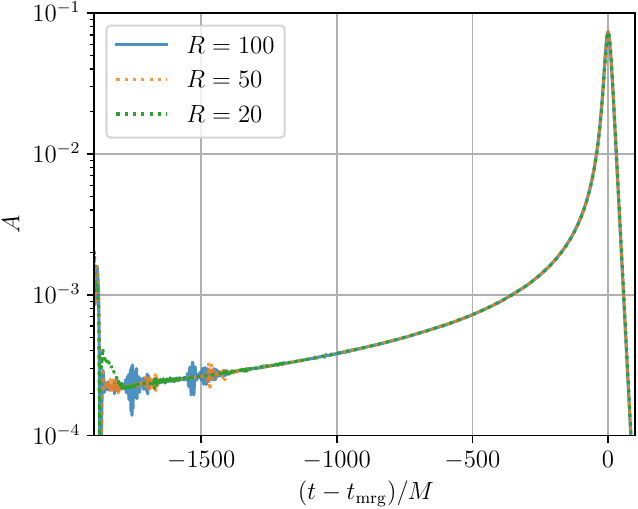}
\caption{%
The amplitude of $\psi^{22}_4$ for various boundary condition extraction
radii.
While at later time of simulation there is no observable difference between
$\psi^{22}_4$ amplitudes, using
smaller radii boundary condition extraction for the \pit{} code exhibit less noise
at the earlier time of simulation.
}
\label{fig:psi4_amp_radial_extraction_effect}
\end{figure}

In order to evaluate the impact of the boundary condition world tube radius on
waveforms, \fig{fig:psi4_amp_radial_extraction_effect} presents the amplitude
of $\mpsi$ obtained using \ac{CCE} for various radii. Upon examining the
waveform at its onset, it becomes apparent that smaller extraction radii yield
smoother results with reduced noise levels.  
However,
it is important to note that although the smallest radius may offer smoother
results, we do not observe a clean convergence with respect to the \pit{}
resolutions.
Consequently, the smallest radius that
demonstrates convergence with respect to the \pit{} resolutions, namely,
$R=50$, see Sec.~(\ref{sec:cce_convergence}), 
is selected for the purposes of this study.

\section{Time domain comparisons with SXS}
\label{app:gra_vs_sxs}
\begin{figure*}[t]
  \centering
	\includegraphics[width=1.0\linewidth,clip=true]{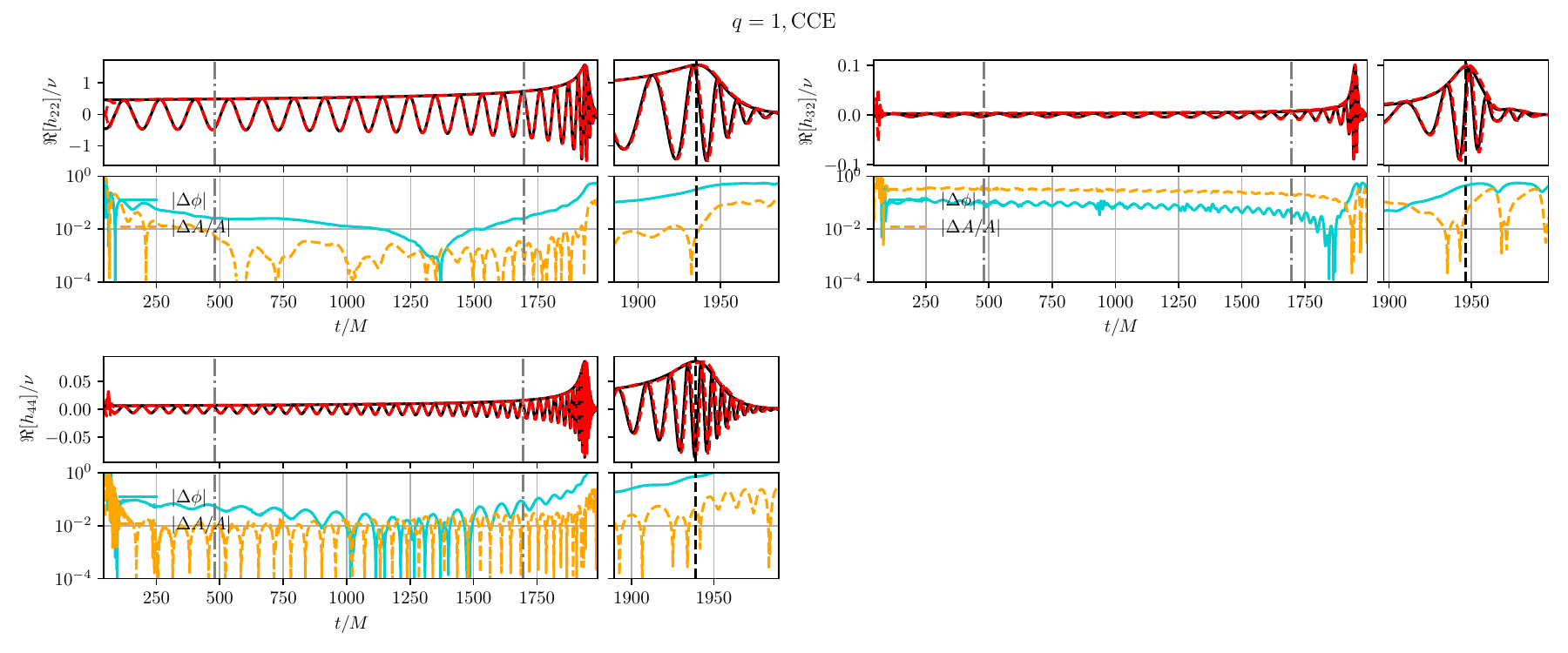}
  \caption{%
  Time domain comparison between the \gra{} (red dashed) and SXS (black) waveforms for the $(\ell, m) = \{(2,2), (3,2), (4,4)\}$ modes and $q=1$.
  Top panels show the real part of the waveform multipoles, while the bottom panels display the phase difference $\Delta\phi = \phi_{\ell m}^{\rm SXS} - \phi_{\ell m}^{\rm GRA}$
  and the amplitude relative difference $\Delta A / A= 1 - A_{\ell m}^{\rm GRA} / A_{\ell m}^{\rm SXS}$. The vertical dashed-doted lines indicate the interval
  used for the alignment, while the vertical dashed lines mark the merger time. We find that while the $(2,2)$ and $(4,4)$ modes exhibit excellent agreement between the two
  waveforms, the $(3,2)$ mode shows larger discrepancies.}
  \label{fig:td_q1}
\end{figure*}

\begin{figure*}[t]
  \centering
	\includegraphics[width=1.0\linewidth,clip=true]{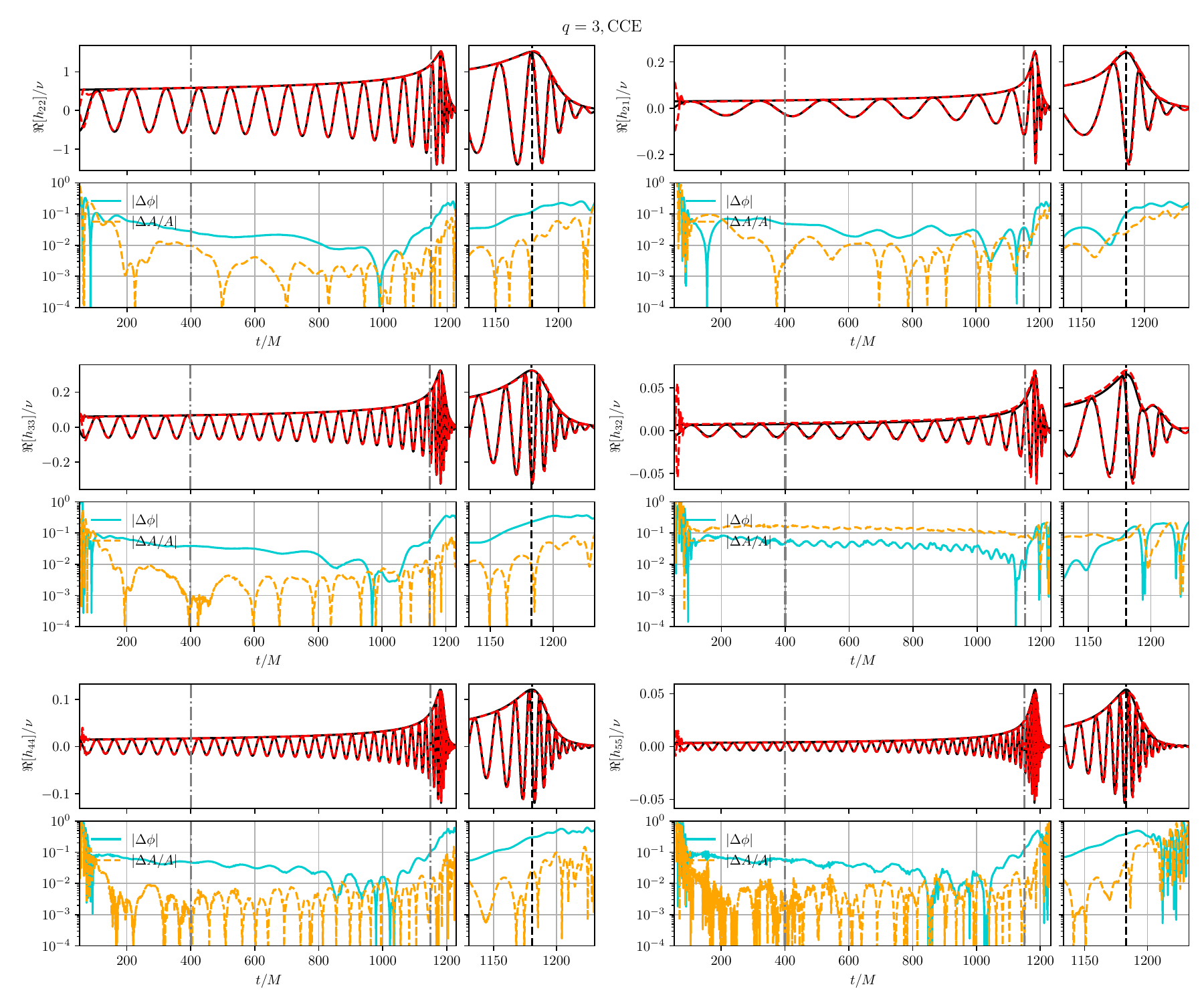}
  \caption{%
  Time domain comparison between the \gra{} (red dashed) and SXS (black) waveforms for the $(\ell, m) = \{(2,2), (2,1), (3,3), (3,2), (4,4), (5,5)\}$ 
  modes and $q=3$. Top panels show the real part of the waveform multipoles, while the bottom panels display the phase difference $\Delta\phi = \phi_{\ell m}^{\rm SXS} - \phi_{\ell m}^{\rm GRA}$
  and the amplitude relative difference $\Delta A / A= 1 - A_{\ell m}^{\rm GRA} / A_{\ell m}^{\rm SXS}$. The vertical dashed-doted lines indicate the interval
  used for the alignment, while the vertical dashed lines mark the merger time. Similarly to the $q=1$ case, modes with $\ell = m$  -- including the $(5,5)$ mode -- are 
  in excellent agreement, while the $(3,2)$ mode shows larger discrepancies.}
  \label{fig:td_q3}
\end{figure*}

\begin{figure*}[t]
  \centering
	\includegraphics[width=1.0\linewidth,clip=true]{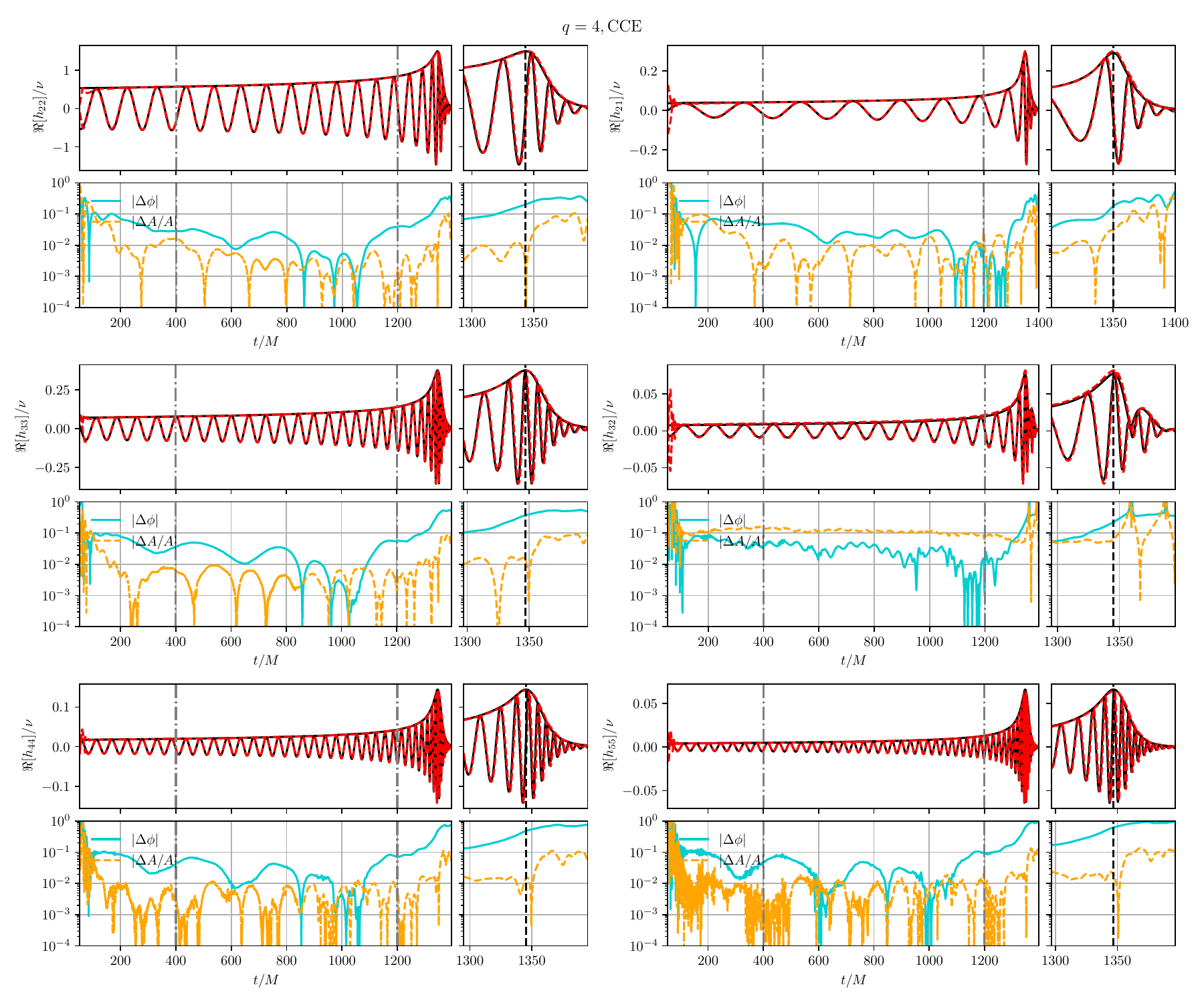}
  \caption{%
  Time domain comparison between the \gra{} (red dashed) and SXS (black) waveforms for the $(\ell, m) = \{(2,2), (2,1), (3,3), (3,2), (4,4), (5,5)\}$
  modes and $q=4$. Top panels show the real part of the waveform multipoles, while the bottom panels display the phase difference $\Delta\phi = \phi_{\ell m}^{\rm SXS} - \phi_{\ell m}^{\rm GRA}$
  and the amplitude relative difference $\Delta A / A= 1 - A_{\ell m}^{\rm GRA} / A_{\ell m}^{\rm SXS}$. The vertical dashed-doted lines indicate the interval
  used for the alignment, while the vertical dashed lines mark the merger time. Despite the
  lower resolution of the GRA waveforms with respect to other mass ratios (320 vs 384),
  the visual agreement and phase differences are comparable to those found with more refinement.}
  \label{fig:td_q4}
\end{figure*}

In this appendix we report time domain comparisons similar to
the one presented in the main text (Fig.~\ref{fig:sxs_td_q2}) for all mass ratios
considered in this work. The results are shown in Figs.~\ref{fig:td_q1}, \ref{fig:td_q3} and
\ref{fig:td_q4}. Quantitative figures such as amplitude and phase differences are comparable
with the ones discussed in the main portion of the text for the $q=2$ system. Phase differences
at merger typically remain below a few portions of a radian, with the notable exception of the $(3,2)$
modes. Notably, this fact remains true also for the $q=4$ simulation, which was run at a lower resolution with respect to
the other systems.


\clearpage

\bibliographystyle{unsrt}

\begin{thebibliography}{100}

\bibitem{GRAthena:BHBH:0001}
David Radice and Alireza Rashti.
\newblock {GRAthena:BHBH:0001}.
\newblock \url{https://doi.org/10.26207/vnq9-za55}, September 2024.

\bibitem{GRAthena:BHBH:0002}
David Radice and Alireza Rashti.
\newblock {GRAthena:BHBH:0002}.
\newblock \url{https://doi.org/10.26207/ykyk-7b04}, September 2024.

\bibitem{GRAthena:BHBH:0003}
David Radice and Alireza Rashti.
\newblock {GRAthena:BHBH:0003}.
\newblock \url{https://doi.org/10.26207/yst7-9m47}, September 2024.

\bibitem{GRAthena:BHBH:0004}
David Radice and Alireza Rashti.
\newblock {GRAthena:BHBH:0004}.
\newblock \url{https://doi.org/10.26207/2rew-2b38}, September 2024.

\bibitem{Pretorius:2005gq}
Frans Pretorius.
\newblock {Evolution of binary black hole spacetimes}.
\newblock {\em Phys. Rev. Lett.}, 95:121101, 2005.

\bibitem{Baker:2006ha}
John~G. Baker, James~R. van Meter, Sean~T. McWilliams, Joan Centrella, and
  Bernard~J. Kelly.
\newblock {Consistency of post-Newtonian waveforms with numerical relativity}.
\newblock {\em Phys. Rev. Lett.}, 99:181101, 2007.

\bibitem{Campanelli:2005dd}
Manuela Campanelli, C.~O. Lousto, P.~Marronetti, and Y.~Zlochower.
\newblock {Accurate evolutions of orbiting black-hole binaries without
  excision}.
\newblock {\em Phys. Rev. Lett.}, 96:111101, 2006.

\bibitem{Hahn:1964a}
Susan~G. {Hahn} and Richard~W. {Lindquist}.
\newblock {The two-body problem in geometrodynamics}.
\newblock {\em Annals of Physics}, 29(2):304--331, September 1964.

\bibitem{Scheel:2006gg}
Mark~A. Scheel, Harald~P. Pfeiffer, Lee Lindblom, Lawrence~E. Kidder, Oliver
  Rinne, and Saul~A. Teukolsky.
\newblock {Solving Einstein's equations with dual coordinate frames}.
\newblock {\em Phys. Rev. D}, 74:104006, 2006.

\bibitem{Sperhake:2006cy}
Ulrich Sperhake.
\newblock {Binary black-hole evolutions of excision and puncture data}.
\newblock {\em Phys. Rev. D}, 76:104015, 2007.

\bibitem{Bruegmann:2006ulg}
Bernd Bruegmann, Jose~A. Gonzalez, Mark Hannam, Sascha Husa, Ulrich Sperhake,
  and Wolfgang Tichy.
\newblock {Calibration of Moving Puncture Simulations}.
\newblock {\em Phys. Rev. D}, 77:024027, 2008.

\bibitem{Brugmann:2008zz}
Bernd Bruegmann, Jose~A. Gonzalez, Mark Hannam, Sascha Husa, Ulrich Sperhake,
  and Wolfgang Tichy.
\newblock {Calibration of Moving Puncture Simulations}.
\newblock {\em Phys. Rev. D}, 77:024027, 2008.

\bibitem{Szilagyi:2009qz}
Bela Szilagyi, Lee Lindblom, and Mark~A. Scheel.
\newblock {Simulations of Binary Black Hole Mergers Using Spectral Methods}.
\newblock {\em Phys. Rev. D}, 80:124010, 2009.

\bibitem{Thierfelder:2011yi}
Marcus Thierfelder, Sebastiano Bernuzzi, and Bernd Bruegmann.
\newblock {Numerical relativity simulations of binary neutron stars}.
\newblock {\em Phys. Rev. D}, 84:044012, 2011.

\bibitem{Loffler:2011ay}
Frank Loffler et~al.
\newblock {The Einstein Toolkit: A Community Computational Infrastructure for
  Relativistic Astrophysics}.
\newblock {\em Class. Quant. Grav.}, 29:115001, 2012.

\bibitem{Babiuc:2010ze}
M.~C. Babiuc, B.~Szilagyi, J.~Winicour, and Y.~Zlochower.
\newblock {A Characteristic Extraction Tool for Gravitational Waveforms}.
\newblock {\em Phys. Rev. D}, 84:044057, 2011.

\bibitem{Hilditch:2015aba}
David Hilditch, Andreas Weyhausen, and Bernd Br\"ugmann.
\newblock {Pseudospectral method for gravitational wave collapse}.
\newblock {\em Phys. Rev. D}, 93(6):063006, 2016.

\bibitem{Bugner:2015gqa}
Marcus Bugner, Tim Dietrich, Sebastiano Bernuzzi, Andreas Weyhausen, and Bernd
  Br\"ugmann.
\newblock {Solving 3D relativistic hydrodynamical problems with weighted
  essentially nonoscillatory discontinuous Galerkin methods}.
\newblock {\em Phys. Rev. D}, 94(8):084004, 2016.

\bibitem{Clough:2015sqa}
Katy Clough, Pau Figueras, Hal Finkel, Markus Kunesch, Eugene~A. Lim, and Saran
  Tunyasuvunakool.
\newblock {GRChombo : Numerical Relativity with Adaptive Mesh Refinement}.
\newblock {\em Class. Quant. Grav.}, 32(24):245011, 2015.

\bibitem{Kidder:2016hev}
Lawrence~E. Kidder et~al.
\newblock {SpECTRE: A Task-based Discontinuous Galerkin Code for Relativistic
  Astrophysics}.
\newblock {\em J. Comput. Phys.}, 335:84--114, 2017.

\bibitem{Daszuta:2021ecf}
Boris Daszuta, Francesco Zappa, William Cook, David Radice, Sebastiano
  Bernuzzi, and Viktoriya Morozova.
\newblock {GR-Athena++: Puncture Evolutions on Vertex-centered Oct-tree
  Adaptive Mesh Refinement}.
\newblock {\em Astrophys. J. Supp.}, 257(2):25, 2021.

\bibitem{Buonanno:1998gg}
A.~Buonanno and T.~Damour.
\newblock {Effective one-body approach to general relativistic two-body
  dynamics}.
\newblock {\em Phys. Rev. D}, 59:084006, 1999.

\bibitem{Buonanno:2000ef}
Alessandra Buonanno and Thibault Damour.
\newblock {Transition from inspiral to plunge in binary black hole
  coalescences}.
\newblock {\em Phys. Rev. D}, 62:064015, 2000.

\bibitem{Ramos-Buades:2021adz}
Antoni Ramos-Buades, Alessandra Buonanno, Mohammed Khalil, and Serguei
  Ossokine.
\newblock {Effective-one-body multipolar waveforms for eccentric binary black
  holes with nonprecessing spins}.
\newblock {\em Phys. Rev. D}, 105(4):044035, 2022.

\bibitem{Pompili:2023tna}
Lorenzo Pompili et~al.
\newblock {Laying the foundation of the effective-one-body waveform models
  SEOBNRv5: Improved accuracy and efficiency for spinning nonprecessing binary
  black holes}.
\newblock {\em Phys. Rev. D}, 108(12):124035, 2023.

\bibitem{Ramos-Buades:2023ehm}
Antoni Ramos-Buades, Alessandra Buonanno, H\'ector Estell\'es, Mohammed Khalil,
  Deyan~P. Mihaylov, Serguei Ossokine, Lorenzo Pompili, and Mahlet Shiferaw.
\newblock {Next generation of accurate and efficient multipolar precessing-spin
  effective-one-body waveforms for binary black holes}.
\newblock {\em Phys. Rev. D}, 108(12):124037, 2023.

\bibitem{Chiaramello:2020ehz}
Danilo Chiaramello and Alessandro Nagar.
\newblock {Faithful analytical effective-one-body waveform model for
  spin-aligned, moderately eccentric, coalescing black hole binaries}.
\newblock {\em Phys. Rev. D}, 101(10):101501, 2020.

\bibitem{Akcay:2020qrj}
Sarp Akcay, Rossella Gamba, and Sebastiano Bernuzzi.
\newblock {Hybrid post-Newtonian effective-one-body scheme for spin-precessing
  compact-binary waveforms up to merger}.
\newblock {\em Phys. Rev. D}, 103(2):024014, 2021.

\bibitem{Gamba:2021ydi}
Rossella Gamba, Sarp Ak\c{c}ay, Sebastiano Bernuzzi, and Jake Williams.
\newblock {Effective-one-body waveforms for precessing coalescing compact
  binaries with post-Newtonian twist}.
\newblock {\em Phys. Rev. D}, 106(2):024020, 2022.

\bibitem{Nagar:2023zxh}
Alessandro Nagar, Piero Rettegno, Rossella Gamba, Simone Albanesi, Angelica
  Albertini, and Sebastiano Bernuzzi.
\newblock {Analytic systematics in next generation of effective-one-body
  gravitational waveform models for future observations}.
\newblock {\em Phys. Rev. D}, 108(12):124018, 2023.

\bibitem{Nagar:2024dzj}
Alessandro Nagar, Rossella Gamba, Piero Rettegno, Veronica Fantini, and
  Sebastiano Bernuzzi.
\newblock {Effective-one-body waveform model for non-circularized, planar,
  coalescing black hole binaries: the importance of radiation reaction}.
\newblock 4 2024.

\bibitem{Nagar:2024oyk}
Alessandro Nagar, Sebastiano Bernuzzi, Danilo Chiaramello, Veronica Fantini,
  Rossella Gamba, Mattia Panzeri, and Piero Rettegno.
\newblock {Effective-one-body waveform model for noncircularized, planar,
  coalescing black hole binaries II: high accuracy by improving logarithmic
  terms in resummations}.
\newblock 7 2024.

\bibitem{Ajith:2007qp}
Parameswaran Ajith et~al.
\newblock {Phenomenological template family for black-hole coalescence
  waveforms}.
\newblock {\em Class. Quant. Grav.}, 24:S689--S700, 2007.

\bibitem{Ajith:2007kx}
P.~Ajith et~al.
\newblock {A Template bank for gravitational waveforms from coalescing binary
  black holes. I. Non-spinning binaries}.
\newblock {\em Phys. Rev. D}, 77:104017, 2008.
\newblock [Erratum: Phys.Rev.D 79, 129901 (2009)].

\bibitem{Ajith:2009bn}
P.~Ajith et~al.
\newblock {Inspiral-merger-ringdown waveforms for black-hole binaries with
  non-precessing spins}.
\newblock {\em Phys. Rev. Lett.}, 106:241101, 2011.

\bibitem{Santamaria:2010yb}
L.~Santamaria et~al.
\newblock {Matching post-Newtonian and numerical relativity waveforms:
  systematic errors and a new phenomenological model for non-precessing black
  hole binaries}.
\newblock {\em Phys. Rev. D}, 82:064016, 2010.

\bibitem{Husa:2015iqa}
Sascha Husa, Sebastian Khan, Mark Hannam, Michael P\"urrer, Frank Ohme, Xisco
  Jim\'enez~Forteza, and Alejandro Boh\'e.
\newblock {Frequency-domain gravitational waves from nonprecessing black-hole
  binaries. I. New numerical waveforms and anatomy of the signal}.
\newblock {\em Phys. Rev. D}, 93(4):044006, 2016.

\bibitem{Khan:2015jqa}
Sebastian Khan, Sascha Husa, Mark Hannam, Frank Ohme, Michael P\"urrer, Xisco
  Jim\'enez~Forteza, and Alejandro Boh\'e.
\newblock {Frequency-domain gravitational waves from nonprecessing black-hole
  binaries. II. A phenomenological model for the advanced detector era}.
\newblock {\em Phys. Rev. D}, 93(4):044007, 2016.

\bibitem{Pratten:2020fqn}
Geraint Pratten, Sascha Husa, Cecilio Garcia-Quiros, Marta Colleoni, Antoni
  Ramos-Buades, Hector Estelles, and Rafel Jaume.
\newblock {Setting the cornerstone for a family of models for gravitational
  waves from compact binaries: The dominant harmonic for nonprecessing
  quasicircular black holes}.
\newblock {\em Phys. Rev. D}, 102(6):064001, 2020.

\bibitem{Estelles:2020osj}
H\'ector Estell\'es, Antoni Ramos-Buades, Sascha Husa, Cecilio
  Garc\'\i{}a-Quir\'os, Marta Colleoni, Le\"\i{}la Haegel, and Rafel Jaume.
\newblock {Phenomenological time domain model for dominant quadrupole
  gravitational wave signal of coalescing binary black holes}.
\newblock {\em Phys. Rev. D}, 103(12):124060, 2021.

\bibitem{Estelles:2021gvs}
H\'ector Estell\'es, Marta Colleoni, Cecilio Garc\'\i{}a-Quir\'os, Sascha Husa,
  David Keitel, Maite Mateu-Lucena, Maria de~Lluc Planas, and Antoni
  Ramos-Buades.
\newblock {New twists in compact binary waveform modeling: A fast time-domain
  model for precession}.
\newblock {\em Phys. Rev. D}, 105(8):084040, 2022.

\bibitem{Hamilton:2021pkf}
Eleanor Hamilton, Lionel London, Jonathan~E. Thompson, Edward Fauchon-Jones,
  Mark Hannam, Chinmay Kalaghatgi, Sebastian Khan, Francesco Pannarale, and
  Alex Vano-Vinuales.
\newblock {Model of gravitational waves from precessing black-hole binaries
  through merger and ringdown}.
\newblock {\em Phys. Rev. D}, 104(12):124027, 2021.

\bibitem{London:2017bcn}
Lionel London, Sebastian Khan, Edward Fauchon-Jones, Cecilio Garc\'\i{}a, Mark
  Hannam, Sascha Husa, Xisco Jim\'enez-Forteza, Chinmay Kalaghatgi, Frank Ohme,
  and Francesco Pannarale.
\newblock {First higher-multipole model of gravitational waves from spinning
  and coalescing black-hole binaries}.
\newblock {\em Phys. Rev. Lett.}, 120(16):161102, 2018.

\bibitem{Garcia-Quiros:2020qpx}
Cecilio Garc\'\i{}a-Quir\'os, Marta Colleoni, Sascha Husa, H\'ector Estell\'es,
  Geraint Pratten, Antoni Ramos-Buades, Maite Mateu-Lucena, and Rafel Jaume.
\newblock {Multimode frequency-domain model for the gravitational wave signal
  from nonprecessing black-hole binaries}.
\newblock {\em Phys. Rev. D}, 102(6):064002, 2020.

\bibitem{Khan:2019kot}
Sebastian Khan, Frank Ohme, Katerina Chatziioannou, and Mark Hannam.
\newblock {Including higher order multipoles in gravitational-wave models for
  precessing binary black holes}.
\newblock {\em Phys. Rev. D}, 101(2):024056, 2020.

\bibitem{Hannam:2013oca}
Mark Hannam, Patricia Schmidt, Alejandro Boh\'e, Le\"\i{}la Haegel, Sascha
  Husa, Frank Ohme, Geraint Pratten, and Michael P\"urrer.
\newblock {Simple Model of Complete Precessing Black-Hole-Binary Gravitational
  Waveforms}.
\newblock {\em Phys. Rev. Lett.}, 113(15):151101, 2014.

\bibitem{Schmidt:2014iyl}
Patricia Schmidt, Frank Ohme, and Mark Hannam.
\newblock {Towards models of gravitational waveforms from generic binaries II:
  Modelling precession effects with a single effective precession parameter}.
\newblock {\em Phys. Rev. D}, 91(2):024043, 2015.

\bibitem{Khan:2018fmp}
Sebastian Khan, Katerina Chatziioannou, Mark Hannam, and Frank Ohme.
\newblock {Phenomenological model for the gravitational-wave signal from
  precessing binary black holes with two-spin effects}.
\newblock {\em Phys. Rev. D}, 100(2):024059, 2019.

\bibitem{Pratten:2020ceb}
Geraint Pratten et~al.
\newblock {Computationally efficient models for the dominant and subdominant
  harmonic modes of precessing binary black holes}.
\newblock {\em Phys. Rev. D}, 103(10):104056, 2021.

\bibitem{Blackman:2014maa}
Jonathan Blackman, Bela Szilagyi, Chad~R. Galley, and Manuel Tiglio.
\newblock {Sparse Representations of Gravitational Waves from Precessing
  Compact Binaries}.
\newblock {\em Phys. Rev. Lett.}, 113(2):021101, 2014.

\bibitem{Blackman:2017dfb}
Jonathan Blackman, Scott~E. Field, Mark~A. Scheel, Chad~R. Galley, Daniel~A.
  Hemberger, Patricia Schmidt, and Rory Smith.
\newblock {A Surrogate Model of Gravitational Waveforms from Numerical
  Relativity Simulations of Precessing Binary Black Hole Mergers}.
\newblock {\em Phys. Rev. D}, 95(10):104023, 2017.

\bibitem{Varma:2018mmi}
Vijay Varma, Scott~E. Field, Mark~A. Scheel, Jonathan Blackman, Lawrence~E.
  Kidder, and Harald~P. Pfeiffer.
\newblock {Surrogate model of hybridized numerical relativity binary black hole
  waveforms}.
\newblock {\em Phys. Rev. D}, 99(6):064045, 2019.

\bibitem{Varma:2019csw}
Vijay Varma, Scott~E. Field, Mark~A. Scheel, Jonathan Blackman, Davide Gerosa,
  Leo~C. Stein, Lawrence~E. Kidder, and Harald~P. Pfeiffer.
\newblock {Surrogate models for precessing binary black hole simulations with
  unequal masses}.
\newblock {\em Phys. Rev. Research.}, 1:033015, 2019.

\bibitem{Williams:2019vub}
Daniel Williams, Ik~Siong Heng, Jonathan Gair, James~A. Clark, and Bhavesh
  Khamesra.
\newblock {Precessing numerical relativity waveform surrogate model for binary
  black holes: A Gaussian process regression approach}.
\newblock {\em Phys. Rev. D}, 101(6):063011, 2020.

\bibitem{Chandra:2020ccy}
Koustav Chandra, V.~Gayathri, Juan~Calderon Bustillo, and Archana Pai.
\newblock {Numerical relativity injection analysis of signals from generically
  spinning intermediate mass black hole binaries in Advanced LIGO data}.
\newblock {\em Phys. Rev. D}, 102(4):044035, 2020.

\bibitem{LIGOScientific:2014oec}
J.~Aasi et~al.
\newblock {The NINJA-2 project: Detecting and characterizing gravitational
  waveforms modelled using numerical binary black hole simulations}.
\newblock {\em Class. Quant. Grav.}, 31:115004, 2014.

\bibitem{Chandra:2021xvs}
Koustav Chandra, Archana Pai, V.~Villa-Ortega, T.~Dent, C.~McIsaac, I.~W.
  Harry, G.~S.~Cabourn Davies, and K.~Soni.
\newblock {Salient features of the optimised PyCBC IMBH search}.
\newblock In {\em {16th Marcel Grossmann Meeting on~Recent Developments in
  Theoretical and Experimental General Relativity, Astrophysics and
  Relativistic Field Theories}}, 10 2021.

\bibitem{LIGOScientific:2014pky}
J.~Aasi et~al.
\newblock {Advanced LIGO}.
\newblock {\em Class. Quant. Grav.}, 32:074001, 2015.

\bibitem{Reitze:2019iox}
David Reitze et~al.
\newblock {Cosmic Explorer: The U.S. Contribution to Gravitational-Wave
  Astronomy beyond LIGO}.
\newblock {\em Bull. Am. Astron. Soc.}, 51(7):035, 2019.

\bibitem{Evans:2023euw}
Matthew Evans et~al.
\newblock {Cosmic Explorer: A Submission to the NSF MPSAC ngGW Subcommittee}.
\newblock 6 2023.

\bibitem{Punturo:2010zz}
M.~Punturo et~al.
\newblock {The Einstein Telescope: A third-generation gravitational wave
  observatory}.
\newblock {\em Class. Quant. Grav.}, 27:194002, 2010.

\bibitem{Maggiore:2019uih}
Michele Maggiore et~al.
\newblock {Science Case for the Einstein Telescope}.
\newblock {\em JCAP}, 03:050, 2020.

\bibitem{amaro2017laser}
Pau Amaro-Seoane, Heather Audley, Stanislav Babak, John Baker, Enrico Barausse,
  Peter Bender, Emanuele Berti, Pierre Binetruy, Michael Born, Daniele
  Bortoluzzi, et~al.
\newblock Laser interferometer space antenna.
\newblock {\em arXiv preprint arXiv:1702.00786}, 2017.

\bibitem{TianQin:2015yph}
Jun Luo et~al.
\newblock {TianQin: a space-borne gravitational wave detector}.
\newblock {\em Class. Quant. Grav.}, 33(3):035010, 2016.

\bibitem{TaijiScientific:2021qgx}
Yue-Liang Wu et~al.
\newblock {China\textquoteright{}s first step towards probing the expanding
  universe and the nature of gravity using a space borne gravitational wave
  antenna}.
\newblock {\em Commun. Phys.}, 4(1):34, 2021.

\bibitem{Kawamura:2006up}
S.~Kawamura et~al.
\newblock {The Japanese space gravitational wave antenna DECIGO}.
\newblock {\em Class. Quant. Grav.}, 23:S125--S132, 2006.

\bibitem{Ajith:2024mie}
Parameswaran Ajith et~al.
\newblock {The Lunar Gravitational-wave Antenna: Mission Studies and Science
  Case}.
\newblock 4 2024.

\bibitem{Lindblom:2008cm}
Lee Lindblom, Benjamin~J. Owen, and Duncan~A. Brown.
\newblock {Model Waveform Accuracy Standards for Gravitational Wave Data
  Analysis}.
\newblock {\em Phys. Rev. D}, 78:124020, 2008.

\bibitem{Kapil:2024zdn}
Veome Kapil, Luca Reali, Roberto Cotesta, and Emanuele Berti.
\newblock {Systematic bias from waveform modeling for binary black hole
  populations in next-generation gravitational wave detectors}.
\newblock {\em Phys. Rev. D}, 109(10):104043, 2024.

\bibitem{Dhani:2024jja}
Arnab Dhani, Sebastian V\"olkel, Alessandra Buonanno, Hector Estelles, Jonathan
  Gair, Harald~P. Pfeiffer, Lorenzo Pompili, and Alexandre Toubiana.
\newblock {Systematic Biases in Estimating the Properties of Black Holes Due to
  Inaccurate Gravitational-Wave Models}.
\newblock 4 2024.

\bibitem{Chandra:2024dhf}
Koustav Chandra.
\newblock {gwforge: A user-friendly package to generate gravitational-wave mock
  data}.
\newblock 7 2024.

\bibitem{Purrer:2019jcp}
Michael P\"urrer and Carl-Johan Haster.
\newblock {Gravitational waveform accuracy requirements for future ground-based
  detectors}.
\newblock {\em Phys. Rev. Res.}, 2(2):023151, 2020.

\bibitem{Jan:2023raq}
Aasim Jan, Deborah Ferguson, Jacob Lange, Deirdre Shoemaker, and Aaron
  Zimmerman.
\newblock {Accuracy limitations of existing numerical relativity waveforms on
  the data analysis of current and future ground-based detectors}.
\newblock {\em Phys. Rev. D}, 110(2):024023, 2024.

\bibitem{Ferguson:2020xnm}
Deborah Ferguson, Karan Jani, Pablo Laguna, and Deirdre Shoemaker.
\newblock {Assessing the readiness of numerical relativity for LISA and 3G
  detectors}.
\newblock {\em Phys. Rev. D}, 104(4):044037, 2021.

\bibitem{Ajith:2012az}
P.~Ajith et~al.
\newblock {The NINJA-2 catalog of hybrid post-Newtonian/numerical-relativity
  waveforms for non-precessing black-hole binaries}.
\newblock {\em Class. Quant. Grav.}, 29:124001, 2012.
\newblock [Erratum: Class.Quant.Grav. 30, 199401 (2013)].

\bibitem{Hinder:2013oqa}
Ian Hinder et~al.
\newblock {Error-analysis and comparison to analytical models of numerical
  waveforms produced by the NRAR Collaboration}.
\newblock {\em Class. Quant. Grav.}, 31:025012, 2014.

\bibitem{Rashti:2023wfe}
Alireza Rashti, Maitraya Bhattacharyya, David Radice, Boris Daszuta, William
  Cook, and Sebastiano Bernuzzi.
\newblock {Adaptive mesh refinement in binary black holes simulations}.
\newblock {\em Class. Quant. Grav.}, 41(9):095001, 2024.

\bibitem{Bernuzzi:2009ex}
Sebastiano Bernuzzi and David Hilditch.
\newblock {Constraint violation in free evolution schemes: Comparing BSSNOK
  with a conformal decomposition of Z4}.
\newblock {\em Phys. Rev. D}, 81:084003, 2010.

\bibitem{Hilditch:2012fp}
David Hilditch, Sebastiano Bernuzzi, Marcus Thierfelder, Zhoujian Cao, Wolfgang
  Tichy, and Bernd Bruegmann.
\newblock {Compact binary evolutions with the Z4c formulation}.
\newblock {\em Phys. Rev. D}, 88:084057, 2013.

\bibitem{Stone2020TheAA}
James~M. Stone, Kengo Tomida, Christopher~J. White, and Kyle~Gerard Felker.
\newblock The athena++ adaptive mesh refinement framework: Design and
  magnetohydrodynamic solvers.
\newblock {\em The Astrophysical Journal Supplement Series}, 249, 2020.

\bibitem{Berger:1984zza}
Marsha~J. Berger and Joseph Oliger.
\newblock {Adaptive Mesh Refinement for Hyperbolic Partial Differential
  Equations}.
\newblock {\em J. Comput. Phys.}, 53:484, 1984.

\bibitem{Ansorg:2004ds}
Marcus Ansorg, Bernd Bruegmann, and Wolfgang Tichy.
\newblock {A Single-domain spectral method for black hole puncture data}.
\newblock {\em Phys. Rev. D}, 70:064011, 2004.

\bibitem{Hannam:2010ec}
Mark Hannam, Sascha Husa, Frank Ohme, Doreen Muller, and Bernd Bruegmann.
\newblock {Simulations of black-hole binaries with unequal masses or
  nonprecessing spins: Accuracy, physical properties, and comparison with
  post-Newtonian results}.
\newblock {\em Phys. Rev. D}, 82:124008, 2010.

\bibitem{Ramos-Buades:2018azo}
Antoni Ramos-Buades, Sascha Husa, and Geraint Pratten.
\newblock {Simple procedures to reduce eccentricity of binary black hole
  simulations}.
\newblock {\em Phys. Rev. D}, 99(2):023003, 2019.

\bibitem{Bishop:1998uk}
Nigel~T. Bishop, Roberto Gomez, Luis Lehner, Bela Szilagyi, Jeffrey Winicour,
  and Richard~A. Isaacson.
\newblock {\em {Cauchy characteristic matching}}, pages 383--408.
\newblock 1 1998.

\bibitem{gourgoulhon_thebook}
Eric Gourgoulhon.
\newblock {\em {3 + 1 Formalism in General Relativity}}.
\newblock Number July. Springer, Berlin New York, 2011.

\bibitem{Reisswig:2010di}
Christian Reisswig and Denis Pollney.
\newblock {Notes on the integration of numerical relativity waveforms}.
\newblock {\em Class. Quant. Grav.}, 28:195015, 2011.

\bibitem{Blanchet:1992br}
Luc Blanchet and Thibault Damour.
\newblock {Hereditary effects in gravitational radiation}.
\newblock {\em Phys. Rev. D}, 46:4304--4319, 1992.

\bibitem{Pollney:2010hs}
Denis Pollney and Christian Reisswig.
\newblock {Gravitational memory in binary black hole mergers}.
\newblock {\em Astrophys. J. Lett.}, 732:L13, 2011.

\bibitem{Mitman:2024uss}
Keefe Mitman et~al.
\newblock {A review of gravitational memory and BMS frame fixing in numerical
  relativity}.
\newblock {\em Class. Quant. Grav.}, 41(22):223001, 2024.

\bibitem{Richardson:1911}
L.~F. Richardson.
\newblock The {Approximate} {Arithmetical} {Solution} by {Finite} {Differences}
  of {Physical} {Problems} {Involving} {Differential} {Equations}, with an
  {Application} to the {Stresses} in a {Masonry} {Dam}.
\newblock {\em Philosophical Transactions of the Royal Society A: Mathematical,
  Physical and Engineering Sciences}, 210(459-470):307--357, January 1911.

\bibitem{alcubierre_book}
Miguel Alcubierre.
\newblock {\em {Introduction to 3+1 Numerical Relativity}}.
\newblock Oxford University Press, 04 2008.

\bibitem{Brill:1959zz}
Dieter~R. Brill.
\newblock {On the positive definite mass of the Bondi-Weber-Wheeler
  time-symmetric gravitational waves}.
\newblock {\em Annals Phys.}, 7:466--483, 1959.

\bibitem{Zhu:2024utz}
Hengrui Zhu, Jacob Fields, Francesco Zappa, David Radice, James Stone, Alireza
  Rashti, William Cook, Sebastiano Bernuzzi, and Boris Daszuta.
\newblock {Performance-Portable Numerical Relativity with AthenaK}.
\newblock 9 2024.

\bibitem{Tichy:2019ouu}
Wolfgang Tichy, Alireza Rashti, Tim Dietrich, Reetika Dudi, and Bernd
  Br\"ugmann.
\newblock {Constructing binary neutron star initial data with high spins, high
  compactnesses, and high mass ratios}.
\newblock {\em Phys. Rev. D}, 100(12):124046, 2019.

\bibitem{Pfeiffer:2007yz}
Harald~P. Pfeiffer, Duncan~A. Brown, Lawrence~E. Kidder, Lee Lindblom, Geoffrey
  Lovelace, and Mark~A. Scheel.
\newblock {Reducing orbital eccentricity in binary black hole simulations}.
\newblock {\em Class. Quant. Grav.}, 24:S59--S82, 2007.

\bibitem{Barkett:2019uae}
Kevin Barkett, Jordan Moxon, Mark~A. Scheel, and B\'ela Szil\'agyi.
\newblock {Spectral Cauchy-Characteristic Extraction of the Gravitational Wave
  News Function}.
\newblock {\em Phys. Rev. D}, 102(2):024004, 2020.

\bibitem{Harry:2017weg}
Ian Harry, Juan Calder\'on~Bustillo, and Alex Nitz.
\newblock {Searching for the full symphony of black hole binary mergers}.
\newblock {\em Phys. Rev. D}, 97(2):023004, 2018.

\bibitem{Chandra:2022ixv}
Koustav Chandra, Juan Calder\'on~Bustillo, Archana Pai, and I.~W. Harry.
\newblock {First gravitational-wave search for intermediate-mass black hole
  mergers with higher-order harmonics}.
\newblock {\em Phys. Rev. D}, 106(12):123003, 2022.

\bibitem{LISA:2017pwj}
Pau Amaro-Seoane et~al.
\newblock {Laser Interferometer Space Antenna}.
\newblock 2 2017.

\bibitem{Vaishnav:2007nm}
Birjoo Vaishnav, Ian Hinder, Frank Herrmann, and Deirdre Shoemaker.
\newblock {Matched filtering of numerical relativity templates of spinning
  binary black holes}.
\newblock {\em Phys. Rev. D}, 76:084020, 2007.

\bibitem{Babak:2021mhe}
Stanislav Babak, Antoine Petiteau, and Martin Hewitson.
\newblock {LISA Sensitivity and SNR Calculations}.
\newblock 8 2021.

\bibitem{Buchman:2012dw}
Luisa~T. Buchman, Harald~P. Pfeiffer, Mark~A. Scheel, and Bela Szilagyi.
\newblock {Simulations of non-equal mass black hole binaries with spectral
  methods}.
\newblock {\em Phys. Rev. D}, 86:084033, 2012.

\bibitem{Blackman:2015pia}
Jonathan Blackman, Scott~E. Field, Chad~R. Galley, B\'ela Szil\'agyi, Mark~A.
  Scheel, Manuel Tiglio, and Daniel~A. Hemberger.
\newblock {Fast and Accurate Prediction of Numerical Relativity Waveforms from
  Binary Black Hole Coalescences Using Surrogate Models}.
\newblock {\em Phys. Rev. Lett.}, 115(12):121102, 2015.

\bibitem{Boyle:2019kee}
Michael Boyle et~al.
\newblock {The SXS Collaboration catalog of binary black hole simulations}.
\newblock {\em Class. Quant. Grav.}, 36(19):195006, 2019.

\bibitem{Albertini:2021tbt}
Angelica Albertini, Alessandro Nagar, Piero Rettegno, Simone Albanesi, and
  Rossella Gamba.
\newblock {Waveforms and fluxes: Towards a self-consistent effective one body
  waveform model for nonprecessing, coalescing black-hole binaries for third
  generation detectors}.
\newblock {\em Phys. Rev. D}, 105(8):084025, 2022.

\bibitem{Ferguson:2023vta}
Deborah Ferguson et~al.
\newblock {Second MAYA Catalog of Binary Black Hole Numerical Relativity
  Waveforms}.
\newblock 9 2023.

\bibitem{Moxon:2021gbv}
Jordan Moxon, Mark~A. Scheel, Saul~A. Teukolsky, Nils Deppe, Nils Fischer,
  Francois H\'ebert, Lawrence~E. Kidder, and William Throwe.
\newblock {SpECTRE Cauchy-characteristic evolution system for rapid, precise
  waveform extraction}.
\newblock {\em Phys. Rev. D}, 107(6):064013, 2023.

\bibitem{Lovelace:2024wra}
Geoffrey Lovelace et~al.
\newblock {Simulating binary black hole mergers using discontinuous Galerkin
  methods}.
\newblock {\em Class. Quant. Grav.}, 42(3):035001, 2025.

\bibitem{Newman:1961qr}
Ezra Newman and Roger Penrose.
\newblock {An Approach to gravitational radiation by a method of spin
  coefficients}.
\newblock {\em J. Math. Phys.}, 3:566--578, 1962.

\bibitem{Lousto:2010qx}
Carlos~O. Lousto, Hiroyuki Nakano, Yosef Zlochower, and Manuela Campanelli.
\newblock {Intermediate-mass-ratio black hole binaries: Intertwining numerical
  and perturbative techniques}.
\newblock {\em Phys. Rev. D}, 82:104057, 2010.

\bibitem{Kiuchi:2017pte}
Kenta Kiuchi, Kyohei Kawaguchi, Koutarou Kyutoku, Yuichiro Sekiguchi, Masaru
  Shibata, and Keisuke Taniguchi.
\newblock {Sub-radian-accuracy gravitational waveforms of coalescing binary
  neutron stars in numerical relativity}.
\newblock {\em Phys. Rev. D}, 96(8):084060, 2017.

\bibitem{Albanesi:2024xus}
Simone Albanesi, Alireza Rashti, Francesco Zappa, Rossella Gamba, William Cook,
  Boris Daszuta, Sebastiano Bernuzzi, Alessandro Nagar, and David Radice.
\newblock {Scattering and dynamical capture of two black holes: synergies
  between numerical and analytical methods}.
\newblock 5 2024.

\end{thebibliography}

\end{document}